\documentclass{aastex63}
\usepackage{lineno}
\usepackage{graphicx}
\usepackage{amsmath}
\usepackage{wasysym}

\newcommand{\ufrac}[2]{{\displaystyle \frac{#1}{\text{\small $#2$}}}}
\newcommand{\fidval}[2]{{\left[\ufrac{#1}{#2}\right]}}
\newcommand{\fidvalpow}[3]{{\fidval{#1}{#2}^{\!#3}}}
\newcommand{\uvec}[1]{{\hat{e}_{#1}}}

\newcommand{\myedit}[1]{{#1}}

\def\orchestra{\textsc{Orchestra}}

\def\micron{\ifmmode {\mu{\rm{m}}}\else $\mu$m\fi}

\def\Msolar{\ifmmode {\rm M_{\odot}}\else $\rm M_{\odot}$\fi}
\def\Rsolar{\ifmmode {\rm R_{\odot}}\else $\rm R_{\odot}$\fi}
\def\Mearth{\ifmmode {\rm M_{\oplus}}\else $\rm M_{\oplus}$\fi}
\def\Rearth{\ifmmode {\rm R_{\oplus}}\else $\rm R_{\oplus}$\fi}

\def\mercsym{{\mercury}}
\def\Mmercury{\ifmmode {\rm M_{_\mercsym}}\else $\rm M_{\mercsym}$\fi}
\def\Rmercury{\ifmmode {\rm R_{\mercsym}}\else ${\rm R_{\mercsym}}$\fi}

\def\rhomerc{\ifmmode {\rho_{\mercsym}}\else $\rho_{\mercsym}$\fi}
\def\cond{\sigma}
\def\mmo{m_\text{mag}}

\def\mmorad{{\xi}_r}
\def\vecmmo{\vec{m}_\text{mag}}
\def\vecmmop{\vec{m}_\text{mag,phys}}
\def\vecmmostar{\vec{m}_\star}
\def\vrel{{v}_\text{rel}}
\def\vecvrel{\vec{v}_\text{rel}}
\def\rcirc{{r}_\text{circ}}
\def\rcircdot{\dot{r}_\text{circ}}

\def\vecBrem{\vec{B}_\text{rem}}

\begin{document}

\title{Magnetic interactions in orbital dynamics}

\author{Benjamin C. Bromley}
\affil{Department of Physics \& Astronomy, University of Utah,
201 JFB, Salt Lake City, UT 84112}
\email{e-mail: bromley@physics.utah.edu}

\author{Scott J. Kenyon}
\affil{Smithsonian Astrophysical Observatory,
60 Garden Street, Cambridge, MA 02138}
\email{e-mail: skenyon@cfa.harvard.edu}

\begin{abstract}
The magnetic field of a host star can impact the orbit of a stellar partner, planet, or asteroid if the orbiting body is itself magnetic or electrically conducting. Here, we focus on the instantaneous magnetic forces on an orbiting body in the limit where the dipole approximation describes its magnetic properties as well as those of its stellar host. A permanent magnet in orbit about a star will be inexorably drawn toward the stellar host if the magnetic force is comparable to gravity due to the steep radial dependence of the dipole-dipole interaction. While magnetic fields in observed systems are much too weak to drive a merger event, we confirm that they may be high enough in some close compact binaries to cause measurable orbital precession. When the orbiting body is a conductor, the stellar field induces a time-varying magnetic dipole moment that leads to the possibility of eccentricity pumping and resonance trapping. The challenge is that the orbiter must be close to the stellar host, so that magnetic interactions must compete with tidal forces and the effects of intense stellar radiation. 
\end{abstract}


\section{Introduction}

Gravity, light, and wind guide the formation of planets around most stars. Yet stellar hosts can also influence material around them through magnetic interactions. A body with a permanent or induced magnetic moment can respond to the local magnetic field tied to a magnetic host star.  While these interactions are weak compared to gravity, cumulative effects of magnetic interactions can change the orbital energy of asteroids \citep{bk2019}, rocky planets \citep{kislyakova2017, kislyakova2018, kislyakova2020, noack2021}, and gas giants \citep{laine2008, laine2012, chang2012, strugarek2017, chyba2021}. Even stellar partners can influence each other magnetically \citep{joss1979, campbell1983, katz1989, katz2017, balazs2021, bourgoin2022}.  

Magnetic interactions can arise between a stellar host and any magnetic or electrically conducting body. A permanently magnetized body will feel a force in a field gradient, while free charges in a small, non-magnetic conductor moving with respect to stellar magnetic field lines experience the Lorentz force. A large, conducting planet forms eddy currents in response to changes in the stellar magnetic field along its orbit \citep[e.g.,][]{giffin2010, nagel2018}; the Lorentz forces on these eddy currents do a small amount of work that can affect orbital energy over time. A conductor in orbit through a static stellar magnetic field can inspiral as a result, with Ohmic losses in the eddy currents draining orbital energy \citep{laine2008, kislyakova2018, bk2019}.  If the star and its magnetic dipole moment are spinning, the orbiting body can migrate outward, tapping the energy of the rotating field. Even when the net time-averaged force is negligible, as when a highly conductive asteroid efficiently generates surface currents to oppose changes in the magnetic flux, the \textit{instantaneous} Lorentz force can be significant \citep{nagel2018}. The magnetic interaction then generates an oscillatory driving force that can impact orbital dynamics.

\citet{goldreich1969} also considered the possibility that Lorentz forces drive current flows in flux tubes between a magnetic primary and a conducting body. In contrast to the induction mechanism just described, in which a conducting planet acts as an AC transformer with an induced magnetic dipole, this unipolar interaction requires a mediating plasma to support the current connecting the primary and the conductor. The original focus of \citet{goldreich1969} on Jupiter-Io as a DC battery has since been broadened to include a wide range of systems \citep[e.g.,][]{lai2012, piro2012}. 

The strength of the stellar magnetic field determines the importance of magnetic interactions compared with gravity. Stars of nearly all spectral types harbor magnetic fields \citep{donati2009}, during the T Tauri phase \citep[e.g.][]{lavail2017, grankin2021}, while on the main sequence \citep{babcock1958, landstreet1992, johns-krull1996, kochukhov2021}, and among compact evolved stars \citep[e.g.,][]{angel1978, valyvin2015, konar2017}. Extreme cases include: Babcock's Star, an Ap star with a $\sim 30$~kG surface field \citep{babcock1960}; white dwarfs with fields around $10^9$~G \citep{caiazzo2021}; and magnetars, with field strengths over $10^{14}$~G \citep{olausen2014}. While dipole field strengths decrease as the inverse distance cubed, extremely magnetic stars produce fields capable of influencing orbital motion at least within a few stellar radii. Around main sequence stars, stellar winds can buoy stellar magnetic fields beyond this distance, supporting a more shallow, $1/r^2$ fall-off \citep[e.g.,][and references therein]{johnstone2012}

An orbiting body's material properties also determine the astrophysical relevance of its magnetic interactions \citep{kislyakova2017, bk2019}.  For example, a conducting body responds to variations in the stellar magnetic field along its orbit by generating eddy currents to oppose changes in magnetic flux. A poor conductor produces only weak currents and barely perturbs the magnetic field. Ohmic losses and long-term changes to the orbit are negligible. At higher conductivity, eddy currents are stronger; significant orbital evolution from Ohmic losses is possible. In the limit of high conductivity, the orbiting body generates surface currents that efficiently repel the flux changes entirely, yet there is little Ohmic dissipation or change in the orbital energy. In realistic astrophysical scenarios, planets consisting of modestly conducting silicates or water may interact magnetically in more interesting ways than bodies made of insulators or superconductors \citep{bk2019}. 

Magnetic interactions have observable astrophysical consequences \citep{joss1979}. For example, a conducting asteroid around a magnetized white dwarf can be drawn inside the Roche radius \citep{bk2019}, where it is eventually accreted by the host star \citep{kb2017wd}, contributing to the `pollution' of the star's atmosphere by metals \citep[e.g.,][]{zuckerman1998, zuckerman2010, kepler2016, farihi2016}. \citet{kislyakova2017} suggested the possibility that Ohmic dissipation can lead to volcanism within rocky planets \citep[see also][]{kislyakova2018, kislyakova2020}, an effect that might enable smaller, close-in bodies to contribute to stellar pollution \citep{bk2019}. More recently, \citet{hogg2021} suggested that the pollution of white dwarfs might be enhanced as small diamagnetic grains on tight orbits interact with the stellar magnetic field. Closer to home, magnetic interactions with Jupiter and its major satellites have served as probes of the moons' internal structure \citep[e.g.,][]{khurana1998, zimmer2000, kivelson2000}.

Here, we concentrate on interactions where the magnetic field is the sole intermediary between a magnetic star and an orbiting partner without a surrounding plasma to sustain a current loop. 
Magnetic forces between bodies are approximated by assigning or deriving dipole fields for the stellar host and its partner. With this idealization, we examine the stability of a stellar binary when each partner has a strong fossil field. We also derive the induced magnetic dipole of planetary companions as they orbit through the stellar magnetic field. Highly conducting planets like Mercury have dipoles that track the local stellar field but are antiparallel to it. A hypothetical planet composed of ferromagnetic material would produce a magnetic dipole moment that lines up with the instantaneous local field. Each of these configurations has a potential impact on orbital dynamics.

This paper begins with a general description of orbital dynamics when dipole-dipole interactions are present alongside gravity (\S2), with the main focus on the stellar magnetic field. Then, in \S3. we discuss the magnetic dipole moments of the stellar or planetary companions of a magnetic star, comparing fossil fields, magnetic material, and conductors. We give examples of specific astrophysical scenarios in \S4, and summarize in \S 5.

\section{Orbital dynamics}\label{sec:magdyn}

\begin{figure}
    \centering
    \includegraphics[width=5in]{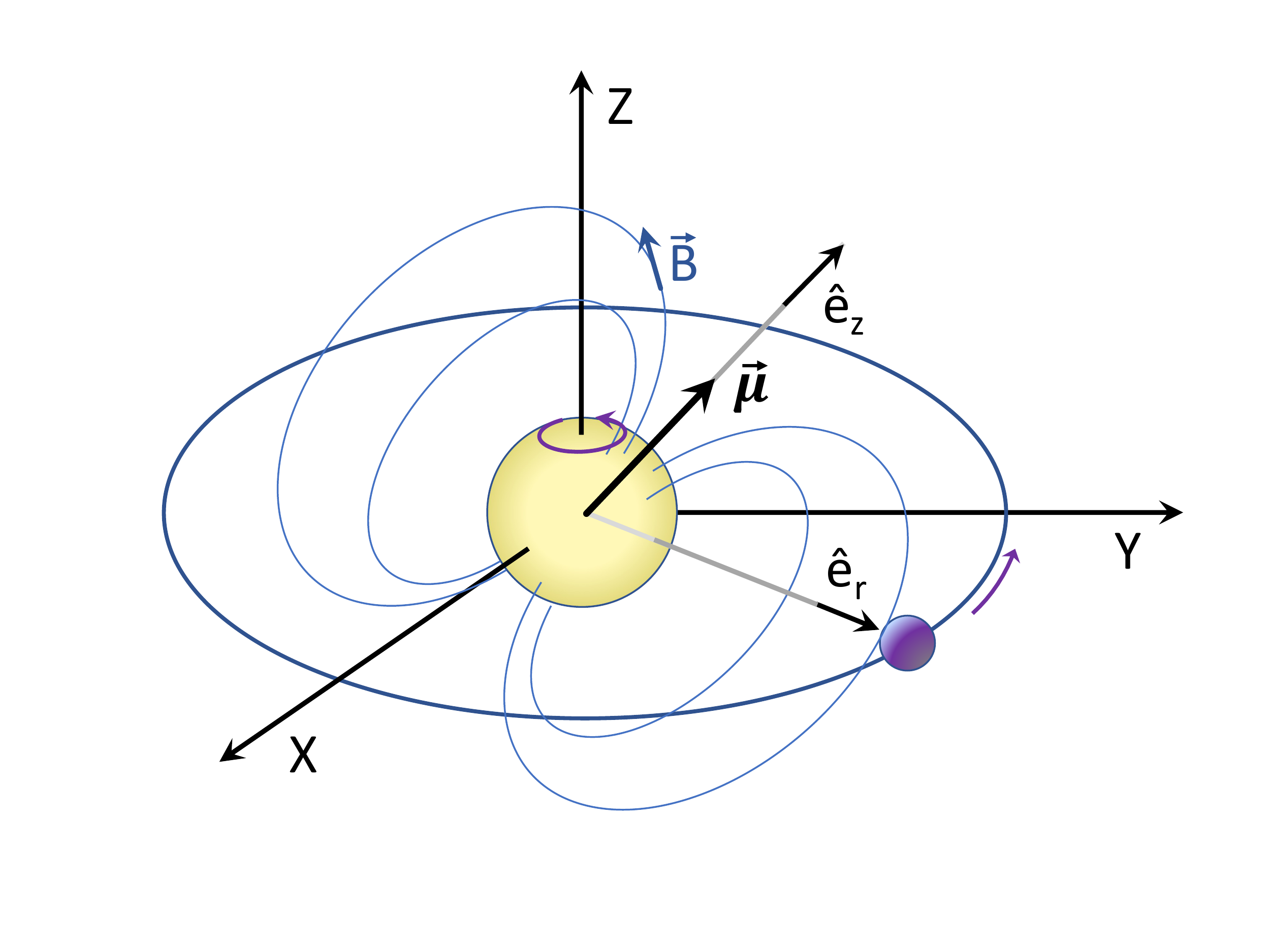}
    \caption{Schematic of a body orbiting a magnetic star. The unit vector $\uvec{z}$ is aligned with the star's magnetic dipole moment $\vec{\mu}$, while the unit vector $\uvec{r}$ gives the direction from the center of the star to the orbiting body. \myedit{Field lines and a sample field vector ($\vec{B}$, in blue) are also indicated to give a sense of the spatial distribution of the magnetic field.} The curved lines with arrows (in purple) show the default sense of stellar spin and the motion of the body. Unless otherwise specified, we assume that the stellar spin and the angular moment of the orbiting body are aligned, as shown.}
    \label{fig:layout}
\end{figure}

To calculate orbital solutions for a body in motion around a magnetized star, we begin by assuming that the stellar magnetic field is a dipole, characterized completely by its magnetic moment,\footnote{Throughout, we adopt the SI system for Maxwell’s equations and derived quantities, following \citet{bidinosti2007}. When providing characteristic values of observed quantities, we use a mix of units to match typical uses in the literature.}
\begin{equation}\label{eq:mmostar}
\vecmmostar = \frac{4\pi B_\star R_\star^3}{\mu_0}\uvec{z},
\end{equation}
where $R_\star$ is the star's radius, $B_\star$ is the surface field strength at the dipolar equator, $\mu_0$ is the permeability of free space, and unit vector $\uvec{z}$ specifies the orientation of the dipole. The instantaneous magnetic field exterior to the star is 
\begin{equation}\label{eq:Bstardipole}
\vec{B}(\vec{r}) = B_\star \frac{R_\star^3}{r^3} 
\left[ 3\uvec{r}(\uvec{z} \cdot \uvec{r})-\uvec{z}\right]
\end{equation}
where $\vec{r}$ is a position relative to the star's center of mass, while $\uvec{r}$ is the unit vector in the direction of $\vec{r}$. When the magnetic moment is not aligned with the star's spin axis, it varies in time as the star rotates; the unit vector $\uvec{z}$ changes accordingly. For simplicity, and unless otherwise stated, we assume throughout that $\vec{r}$ refers to the orbiting body's location and that it lies in a plane perpendicular to the star's spin axis. Figure~\ref{fig:layout} illustrates this geometry.

A magnetized or conducting body orbiting through the stellar magnetic field feels a magnetic force. We work in the limit where the body's size is small compared with the spatial variations in the stellar magnetic field. If the `body' is a stellar companion, then our assumption is that the stars are separated by a distance that is large compared with the secondary's radius. As with the host star, we assume that the body's magnetic properties are encoded by its magnetic moment, $\vecmmo$, the dipole moment of field that the body itself generates, either because of its own material properties or in response to the stellar field. Then, the instantaneous magnetic force that the body feels as it orbits the star is
\begin{equation}\label{eq:magforce}
    \vec{F}_B = \vec{\nabla} \left(\vecmmo \cdot \vec{B}\right)
 = \vecmmo \cdot \vec{\nabla} \vec{B}
\end{equation}
where the right-most equation treats the magnetic moment as independent of position, and $\vec{\nabla} \vec{B}$, with $(\vec{\nabla} \vec{B})_{ij} \equiv dB_i/dx_j$, is the Jacobian of the stellar magnetic field. In a standard $(x,y,z)$ rectilinear coordinate system with the $z$-axis tied to the stellar dipole moment,
\begin{equation}
\vec{B}(\vec{r}) = B_\star \frac{R_\star^3}{r^5} [3 x z \uvec{x} + 3 y z \uvec{y} + (3z^2-r^2)\uvec{z}] 
\end{equation}
\myedit{where $\uvec{i}$ are unit vectors so that position vector $\vec{r} = x\uvec{x}+y\uvec{y}+z\uvec{z}$. Then, the Jacobian is} 
\begin{equation}\label{eq:Jacobian}
    \vec{\nabla}\vec{B} = 3 B_\star \frac{R_\star^3}{r^7}
    \left[
    \begin{array}{ccc}
    z r^2 -5 x^2 z & -5 x y z & x r^2 -5 x z^2  \\
    -5 x y z & z r^2 -5 y^2 z & y r^2 - 5 y z^2  \\
    x r^2 - 5 x z^2 & y r^2 - 5 y z^2 & 3 z r^2 - 5 z^3
    \end{array}
    \right].
\end{equation}
 If we can determine the magnetic moment of an orbiting body, we can calculate the instantaneous magnetic force on it, and thus derive an orbit solution. Because of our choice of simple orbital geometries, as well as magnetic moments of orbiters that are induced by the stellar field, we focus on this center-of-mass force and do not consider magnetic torques.
 
The magnetic force in Equation~(\ref{eq:magforce}) has a qualitatively simple description for an orbit in the plane perpendicular to a fixed stellar magnetic moment. Then, with $z=0$, the Jacobian reduces to
\begin{equation}\label{eq:Jacobianeq}
    \vec{\nabla}\vec{B} = 3 B_\star \frac{R_\star^3}{r^4} \left(\uvec{r}\uvec{z} + \uvec{z}\uvec{r} \right),
\end{equation}
\myedit{where dyadic notation is used to indicate the outer product of these unit vectors.}
If the magnetic moment of the orbiting body is parallel or antiparallel to the stellar dipole, it feels a force in the radial direction from a gradient in the strength of the stellar field.  If the orbiting body's magnetic moment is radially directed, it experiences a vertical force produced by a vertical field gradient associated with the curved field lines of the stellar dipole. 

\subsection{Orbital stability}\label{subsec:magrmsco}

To derive how magnetic forces may impact orbits around a magnetic star, we assume an idealized configuration in which the orbiting body has a magnetic moment that is antiparallel to the stellar dipole and moves in the plane perpendicular to its magnetic moment vector. Then, the force on the body is purely radial, corresponding to a specific potential energy 
\begin{equation}\label{eq:Umag}
    U_\text{mag} =  - \frac{3}{3+\gamma}
    \frac{\mmorad B_\star R_\star^3 (M_s+M_\star)}{M_s M_\star r^{3+\gamma}},
\end{equation}
where $M_s$ is the mass of the orbiter, and we have expressed the magnetic moment in the form
\begin{equation}
    \mmo = \mmorad r^{-\gamma};
\end{equation}
the power-law index $\gamma = 0$ corresponds to an orbiting body with a fixed magnetic moment, while $\gamma = 3$ applies to a planet or asteroid whose magnetic moment is induced by the stellar magnetic field (\S\ref{sec:magdyn} and Appendix \ref{appx:powah}). By including both the orbiter mass and the mass of the star, this potential could apply when the `orbiting body' is a stellar companion, perhaps the secondary partner in a magnetized binary system. The effective radial potential for reduced two-body motion is
\begin{equation}\label{eq:Ueff}
    U_\text{eff} = \frac{L^2}{2 r^2} - \frac{G (M_s+M_\star)}{r} + U_\text{mag},
\end{equation}
where $L$ is the specific angular momentum. The first derivative of $U_\text{eff}$ with respect to $r$,
\begin{equation}\label{eq:dUeff}
    dU_\text{eff}/dr = -\frac{L^2}{r^3} + \frac{3 \mmorad B_\star R_\star^3 (M_s+M_\star)}{M_s M_\star r^{4+\gamma}} + \frac{G (M_s+M_\star)}{r^2},
\end{equation}
 has roots that correspond to circular orbits. The stability of these orbits is determined by the sign of the second derivative,
\begin{equation}
d^2U_\text{eff}/dr^2 = 3\frac{L^2}{r^4} - \frac{3 (4+\gamma) \mmorad B_\star R_\star^3 (M_s+M_\star)}{M_s M_\star r^{5+\gamma}} - 2\frac{G (M_s+M_\star)}{r^3};
\end{equation}
if this second derivative is negative, the orbit is unstable and will lead to inspiral. Setting both first and second derivatives of  $U_\text{eff}$ to zero, we solve for the orbital distance that delimits stable and unstable circular orbits. This magnetic `minimum stable circular orbit' is
\begin{equation}\label{eq:stab}
r_\text{msco} = \left[\frac{3(1+\gamma)\mmorad B_\star R_\star^3}{GM_s M_\star}\right]^{1/(2+\gamma)}
\end{equation}
Unsurprisingly, it will turn out that in astrophysical systems, the orbital radius $r_\text{msco}$ formally takes on values that are much less than the physical radii of magnetic stars, so that no such minimum stable orbits exist around these objects (\S\ref{sec:apex}). Magnetically driven mergers are implausible.


\subsection{Orbit precession}\label{subsec:orbsecular}

Magnetic forces  on an asteroid, planet or stellar companion from a magnetic star are weak compared with gravity (\S\ref{sec:apex}). The instantaneous force in Equation~(\ref{eq:magforce}) is a perturbation that has only a small impact on a body's orbit on dynamical time scales. However, it can lead to orbital precession and, for conducting bodies, a steady, long-term loss of energy that drives orbital evolution away from an orbit that corotates with the stellar magnetic field. Here, we assume the stellar field is fixed in a frame tied to the star's rotation, although it need not be aligned with the spin angular momentum vector. \citet{bourgoin2022} provide a more general analysis when both the orbiting body and the stellar host have fixed magnetic dipole moments.

Despite the comparative weakness of magnetic forces in orbital dynamics, they contribute to secular evolution. For example, the time-averaged potential associated with the magnetic interaction (e.g., Eq~(\ref{eq:Umag})), yields apsidal and nodal precession rates, respectively, 
\begin{gather}
    \label{eq:apserate}
    \dot{\varpi}  \approx -\left.\left[ \frac{1}{2\Omega r^2}\frac{d}{dr}\left(r^2\frac{dU_\text{mag}}{dr}\right)\right.\right|_{r=\rcirc,z=0}
    \\
    \label{eq:noderate}
    \dot{\ascnode} \approx \left.\left[\frac{1}{2\Omega z}
    \frac{dU_\text{mag}}{dz}\right.\right|_{r=\rcirc,z=0}.
\end{gather}
where $\rcirc$ is the semimajor axis of the orbit and $\Omega$ is the orbital frequency. When the magnetic dipoles are antiparallel to each other and oriented nearly perpendicular to the orbital plane, we find
\begin{gather}
    \dot{\varpi} \approx \frac{3 (2+\gamma)}{2\sqrt{G}}
    \frac{\mmorad B_\star R_\star^3 \sqrt{M_s+M_\star}}{M_s M_\star r^{7/2+\gamma}}
    \\
   \dot{\ascnode} \approx -\frac{3}{\sqrt{G}}
    \frac{\mmorad B_\star R_\star^3 \sqrt{M_s+M_\star}}{M_s M_\star r^{7/2+\gamma}}.
\end{gather}
When the orbiter's magnetic moment is fixed ($\gamma = 0$), the apsidal and nodal precession rates are equal in magnitude in the limit of small eccentricity \citep[cf.][]{bourgoin2022}. If the orbiter's magnetic moment derives from magnetization or induction in the stellar field ($\gamma = 3$), the nodal precession rate is slower than the apsidal precession rate. 

\subsection{Loss or gain of orbital energy}

If the orbiting body is a conductor, then magnetic interactions can also drive changes to the orbital energy as reflected by its semimajor axis. Induced currents in a conducting asteroid or planet will slowly but steadily dissipate energy from the system through Ohmic heating. Around a non-rotating star with a static magnetic field, the conductor will steadily lose orbital energy and inspiral toward the star; if the star and its magnetic field rotate faster and in the same sense as the conductor, the conductor's orbit will tend to push outward from the star \citep[e.g.,][see also \citealt{bk2019}]{laine2008, laine2012, kislyakova2018, kislyakova2020}. 

The secular change in the orbital energy of a conductor around a magnetic star is determined by the time-averaged mechanical power,
\begin{equation}\label{eq:powahmech}
    P_\text{mech} =  -\left< (\vecmmo\cdot\vec{\nabla}\vec{B})\cdot \vec{v}\right>,
\end{equation}
where $\vec{v}$ is the conductor's velocity as measured in a reference frame tied to the star's magnetic dipole moment. While the mechanical power accounts for the loss of kinetic energy in this frame, the conducting body will gain or lose orbital energy. Equation~(\ref{eq:powahmech}) corresponds to energy that is converted to heat through Ohmic dissipation of currents induced by the stellar magnetic field.  Thus, we can estimate the rate of change in orbital energy directly from the Ohmic heating rate,
\begin{equation}\label{eq:powahJ}
    P_\Omega = \frac{1}{2} \int_{V} dV \frac{|\vec{J}|^2}{\cond} 
\end{equation}
where $|\vec{J}|$ is the maximum amplitude of the current density, $\cond$ is the conductivity, and the integral is over the volume of the conductor.  

Alternatively, the average power dissipated through Ohmic heating may be determined directly from the conductor's magnetic moment. For example, if the stellar magnetic field (or one component of it) has an amplitude of $|B|$, and the frequency of field variation experienced by the conductor from orbital motion and stellar rotation is $\omega$, then the power loss is  
\begin{equation}\label{eq:powahmmo}
P_\Omega =  \frac{1}{2} \omega |B| \Im\left\{\mmo\right\},
\end{equation}
where the complex magnetic moment magnitude keeps track of the phase difference between the induced currents and the field oscillations (Appendix \ref{appx:powah}). A purely real-valued magnetic moment is in phase with the field oscillations; the average magnetic force then vanishes. A purely imaginary magnetic moment is 90$^\circ$ out of phase with the field; then, the magnetic force acts only in one direction, either opposing the orbital motion or aligned with it. In this formalism, the magnetic force is just the real part of Equation~(\ref{eq:magforce}).  Following \citet{laine2008}, the sign of $\omega$ determines whether the body gains or loses orbital energy. For example, if the orbiting body overtakes slowly corotating stellar magnetic field lines, $\omega>0$, which generally leads to a loss of orbital energy. Rapidly corotating field lines sweep up the body, and $\omega<0$. 

Equation~(\ref{eq:powahmmo}) is essentially the time derivative of the dipole's magnetic energy corresponding to a mode oscillating at frequency $\omega$. In the context of orbital dynamics, $\omega$ is the synodic frequency of motion as the conducting body orbits through the stellar host's (possibly) rotating magnetic field. Harmonics of this frequency may also contribute to the total power loss.

Orbital energy changes from Ohmic heating translates to a rate of change in orbital distance,
\begin{equation}\label{eq:dadt}
    \frac{\rcircdot}{\rcirc} \approx \frac{2 \rcirc P_\Omega}{GM_s M_\star}.
\end{equation}
where the power $P_\Omega$ is negative if a body is plowing through a static or slowly rotating magnetic field, and positive if orbital energy is gained at the expense of the magnetic field's rotation (the Ohmic loss occurs in that rotating frame). Eccentricity can also change; energy is transferred between the field and the conductor most strongly at periastron, leading to circularization or eccentricity pumping depending on whether the orbital period is longer than the star's rotation period.

Non-conducting bodies, even if strongly magnetized, do not experience this type of orbital energy exchange. There are interesting potential dynamical effects nonetheless. We explore one example next.

\subsection{Resonance trapping}

As a final example of the impact of magnetic interactions on orbiting bodies, we note that the orbiter experiences an oscillatory driving force depending on the orbital motion relative to the rotation of the star. The frequency of the field variation in the orbit frame is $\omega \equiv \Omega_\star - \Omega$, where $\Omega_\star$ is the stellar spin rate and $\Omega$ is the orbital frequency; for a planet or asteroid with an induced magnetic moment in a tidally locked frame, the driving force has a frequency of $2\omega$. A resonance occurs when
\begin{equation}
\Omega_\star \approx \left\{
\begin{array}{cc}
   (1-1/2n)\Omega  & \ \ (\Omega > \Omega_\star) \\
    (1+n/2)\Omega & \ \ (\Omega  \leq \Omega_\star)
\end{array} 
\right.
\ \ \ \ \ \ \ (n = 1, 2, 3,...).
\end{equation}
The precise resonance condition will be affected by apsidal precession (Eq~(\ref{eq:apserate})). When an object experiences slow changes in orbital elements, specifically semimajor axis, it can get trapped in a resonance, as in planetary systems \citep[e.g.,][]{malhotra1996, wyatt2003} or galactic dynamics \citep[e.g.,][]{moreno2015}. While the trapping is weak, it is plausible, as we demonstrate below for the case of a slowly varying stellar rotation period (\S\ref{sec:apex}). The phenomenon leads to changes in semimajor axis, but also eccentricity pumping and the possibility of capture by the star.

All of the phenomena just described depend on the nature of the magnetic moments generated in the asteroids, planets or stellar companions of a magnetic star. We consider this topic next.

\section{Magnetic moments}

Magnetic interactions between a star and an orbiting companion depend on the star's magnetic field and the magnetic moment of the orbiting companion. In one simple scenario, the orbiter has a fixed, `fossil' field with a magnetic moment that is independent of the orbit. Alternatively, the star's field can induce magnetization (remanence) if the companion is composed of magnetic material. Finally, temporal variations in the local magnetic field from orbital motion and/or stellar rotation induce free currents that generate a magnetic moment in the companion. In this section, we examine all of these cases.

\subsection{An orbiting magnetic body}\label{subsec:orbmag}

The orbiting body's magnetic moment encodes material properties, its fossil magnetization, and its response to the stellar magnetic field. In one simple case, when the orbiting body is a fixed, permanent magnet, the magnetic moment, $\vecmmo$, is constant in time. If the orbiter is a companion star with a fossil field, its magnetic moment may be specified as in Equation~(\ref{eq:mmostar}). Other sources of the magnetic field are possible, too; younger stars in particular may have strong dynamos, e.g., \citealt{johnstone2021}, and planetary fields also may have a dynamical origin \citep[e.g.][]{christensen2010}. If the orbiter behaves as a permanent or hard magnet composed of ferromagnetic material with high coercivity, then the magnetic moment has a strength of 
\begin{equation}
    \vecmmo = \frac{4\pi R^3}{3\mu_0}\vecBrem  \ \ \ (\text{permanent}),
\end{equation}
where $R$ is the magnet's radius, and $\vecBrem$ is the ``remanant'' (or residual) field. For ferrite, the remanence can reach about 3000~G. For comparison, a magnetic white dwarf has the equivalent of a residual field with a strength as high as $10^9$~G. 

When the body is a ``magnetizable'' low-conductivity, soft magnet, its magnetic moment depends on the local value of the stellar magnetic field. In a linear material, the response to an external stellar field is characterized by its magnetic permeability, $\mu$. If the stellar field penetrates into the bulk material unperturbed by induced currents, then the magnetic moment is
\begin{equation}\label{eq:vecmmomagnetized}
\vecmmo = - 4 \pi \frac{B_\star}{\mu_0} \frac{R_\star^3}{r^3} R^3 \frac{\mu-\mu_0}{\mu + 2\mu_0}\uvec{z}  \ \ \ \ (\text{magnetizable}),
\end{equation}
where the $z$-axis aligns with the \textit{stellar} dipole. The physics is familiar from magnetostatics; bathed in an external field, a magnetic body will generate its own field in the same direction as the external one. This field is associated with bound currents that conceptually represent the underlying molecular or sub-atomic response to the external field.

In these examples, the magnetic moment of an orbiting body is either specified independently of the stellar magnetic field,\ or is derived from the local value of that field. In either case, instantaneous forces are straightforward to calculate in an orbit integration code. When the orbiter is also a conductor, its magnetic moment can be more complicated to derive. We consider this situation next.

\subsection{An orbiting conductor}\label{subsec:orbcond}

The magnetic moment of a conducting body is generated by currents induced as the orbiting body experiences temporal variations in the stellar magnetic field. These variations come from orbital motion or rotation of the stellar magnetic moment, or a combination of both. Toward obtaining orbit solutions, we typically assume that the force of gravity dominates the orbital dynamics. Then, osculating Keplerian orbital elements provide the conductor's trajectory through the stellar magnetic field; the field variations it experiences along its orbit and Faraday's law of induction determine the induced currents and the magnetic moment associated with them. 

To derive the magnetic moment, we assume that this body is perfectly spherical with radius $R$, and that its bulk properties, including mass density $\rho$, conductivity $\cond$, and magnetic permeability $\mu$, vary in the radial direction only. As this sphere orbits the star, the stellar magnetic field it travels through varies in time $t$; we focus here on a single oscillatory mode with frequency $\omega$, so that the magnetic field at the orbiter's location is $\vec{B}_0 \exp(-i\omega t)$, where $\vec{B}_0$ is a constant vector that is approximately uniform over the volume of the conductor. We work in a regime where $\omega$ is small enough that the wavelength of light propagating through the sphere at this frequency is larger than the sphere itself. 

The induced current density and the stellar magnetic field are described by Faraday's law and Amp\`{e}re's law, 
\begin{equation}\label{eq:faradayampere}
    \vec{\nabla}\times\frac{\vec{J}}{\cond} = -i\omega\vec{B}, \ \ \ \ \ \vec{\nabla}\times\frac{\vec{B}}{\mu} = \vec{J}, 
\end{equation}
respectively, where we have adopted Ohm's law to write the electric field in Faraday's law in terms of current density ($\vec{E} = \vec{J}/\cond$). These two equations, combined with the curl of either one, along with boundary conditions at the conducting surface and a regularity condition at the conductor's center, lead to a solution for the current density and the magnetic field.  Appendix~\ref{appx:powah} provides details \citep[see][for example]{joss1979, bidinosti2007, laine2008, giffin2010, chang2012, kislyakova2017, bk2019}.

The solutions for $\vec{J}$ show trends that depend on the magnetic Reynolds number, defined as
\begin{equation}\label{eq:Rm}
    R_m \equiv 2 R^2/\delta^2 = \mu\cond\omega R^2,  
\end{equation}    
where $\delta$ is the 'skin depth', which characterizes how deep the stellar magnetic field permeates into a spherical conductor. If $R_m$ is much less than unity, so that $\delta\gg R$, then the conductor is bathing in an unperturbed external field; weak eddy currents form throughout. If $R_m$ is large compared with $\mu^2/\mu_0^2$, the magnetic field can only penetrate into a thin layer of depth $\delta\ll R$ on the conductor's surface, where strong currents are generated (the 'skin effect'). In a magnetic conductor, $\mu/\mu_0 > 1$, when $1 \lesssim R_m \lesssim \mu^2/\mu_0^2$, induced free currents are confined to the surface of the conductor, but the magnetic field penetrates into the sphere's interior, supported by bound currents \citep{bk2019}.

\begin{figure}
\centerline{\includegraphics{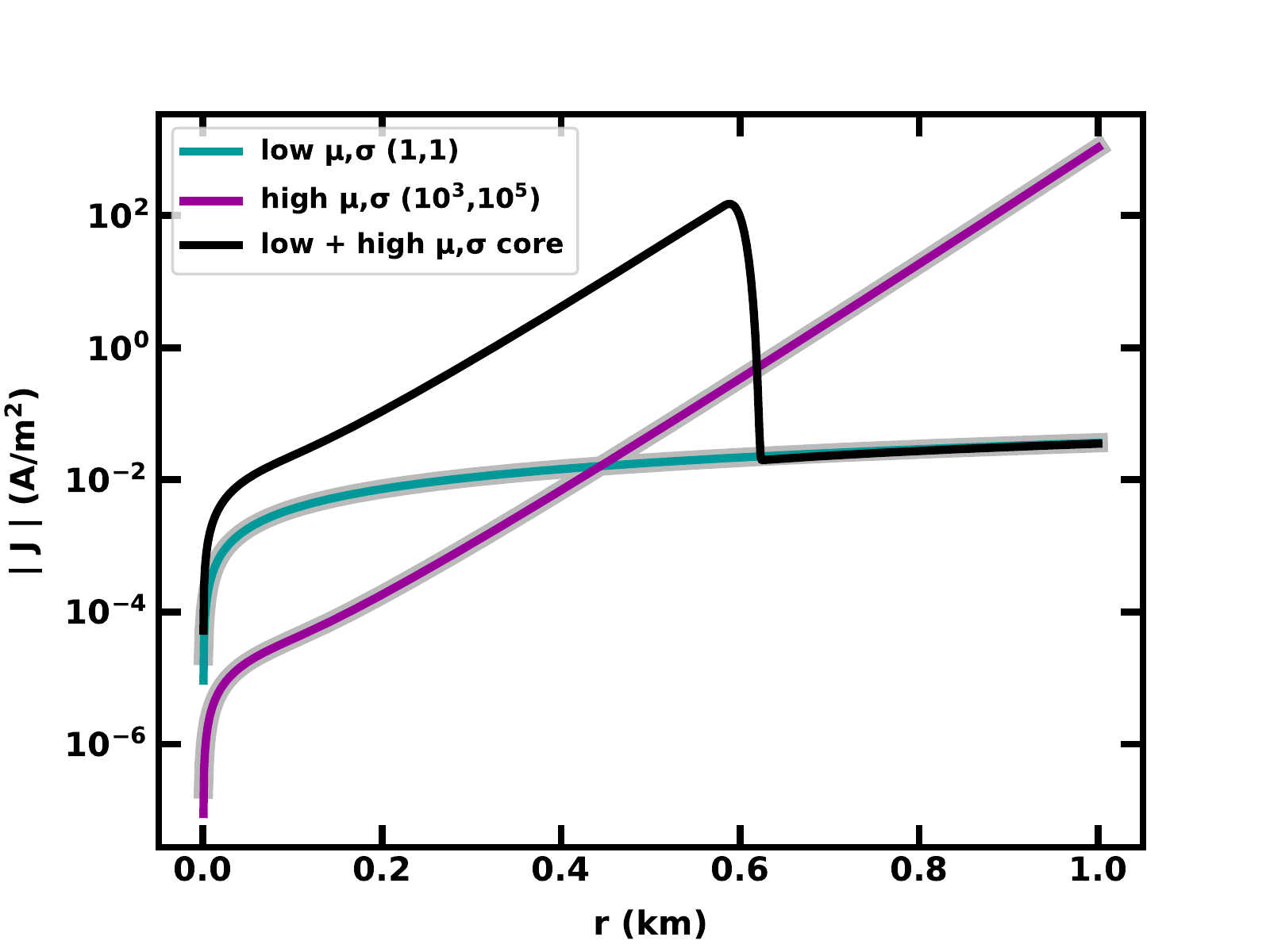}}
    \caption{The peak magnitude of the current density as a function of radius inside kilometer-sized spheres of various materials as determined from numerical calculations. The blue-green curve corresponds to non-magnetic, low-conductivity material in a $|B_0|=1$~T field varying sinusoidally on a 10-day cycle. A sphere made of magnetic, high-conductivity material is indicated in purple. These curves are superimposed on light gray curves from the theoretical predictions of \citet{bidinosti2007}. The legend shows the permeability in units of $\mu_0$ and the conductivity $\sigma$ is in units of S/m. The remaining curve, in black, corresponds to a sphere with an \myedit{outer shell} of the low conductivity material surrounding an inner core of high conductivity and magnetic permeability. To avoid numerical errors, we smoothly transition between high and low values of $\mu$ and $\sigma$ in a radial zone of 0.1~km in thickness that separates these two materials.}
    \label{fig:sphercond}
\end{figure}

Figure~\ref{fig:sphercond} provides an illustration. It shows examples of numerical calculations of the current density for homogeneous conductors, as well as for a sphere with a low-conductivity shell around a highly conductive core. 

The magnetic moment of a conducting body follows directly from the current density:
\begin{equation}\label{eq:vecmmo}
     \vecmmo = \int dV \left[\frac{1}{2}\vec{r}\times \vec{J} 
     + \frac{(\mu-\mu_0)}{\mu} \vec{B} 
     \right]
\end{equation}
where $\vec{J}$ is the current density and $\vec{B} \sim \vec{\nabla}\times \vec{J}/\sigma$ is the magnetic field within the conductor. For a homogeneous medium, \citet{bidinosti2007} give
\begin{equation}
    \mmo = \frac{2\pi R^3 B_0}{\mu_0}\frac{2(\mu-\mu_0)j_0(kR) + (2\mu+\mu_0)j_2(kR)}{(\mu+2\mu_0)j_0(kR)+(\mu-\mu_0)j_2(kR)},
\end{equation}
where $k = \sqrt{i\mu\cond\omega}$ while $j_0$ and $j_2$ are the zeroth- and second-order spherical Bessel functions. In the limits of low and high magnetic Reynolds number, the magnetic moment is
\begin{equation}\label{eq:mmolim}
   |\vecmmo| = 
   \begin{cases}
    \frac{4\pi B_0 R^3}{\mu+2\mu_0} \left|\frac{\mu-\mu_0}{\mu_0} + \frac{3i\mu^2\cond\omega R^2}{10(\mu + 2\mu_0)} \right| & 
(R_m \ll 10)
    \\
    \frac{2\pi B_0 R^3}{\mu_0}  &  (R_m \gg 10 \mu^2/\mu_0^2).
   \end{cases}
\end{equation}
while the magnetic force (Eq.~\ref{eq:magforce}), formally becomes
\begin{equation}
    \vec{F}_B = \vec{B}_0 \cdot \left(\vec{\nabla}\vec{B}\right) \times
   \begin{cases}
    {2\pi i \cond\omega R^5}/{15} & 
(R_m \ll 10 \ \text{and} \ \mu\approx\mu_0)
    \\
    -{2\pi R^3}/{\mu_0}  &  (R_m \gg 10 \mu^2/\mu_0^2).
   \end{cases}
\end{equation}
where the upper equation applies to non-magnetic material.   The transition between these low and high Reynolds number regimes occurs near $R_m \approx 10$ for non-magnetic material, while for magnetic material, there is an intermediate regime that extends from $R_m \sim 10$ to $R_m \sim10 \mu^2/\mu_0^2$, for which the magnet moment originates from a mixture of free and bound currents. 

The power, likewise, has characteristic behavior in each of these regimes. From Equation~(\ref{eq:powahmmo}), 
\begin{equation} 
P_\Omega \approx \begin{cases}
3 \pi B_0^2\{\mu^2 \cond \omega^2 R^5/{5 (\mu+2\mu_0)^2} & R_m \ll 10 \\
2 \pi B_0^2 \sqrt{\cond\omega^3/\mu} R^4 & 10 \lesssim R_m \lesssim 10 \mu^2/\mu_0^2 \\ 
3 \pi B_0^2 \sqrt{\mu\omega/2\cond} R^2 & R_m \gg 10 \mu^2/\mu_0^2.
\end{cases}
\end{equation}
The top and bottom expressions are exact in the limits shown, while the middle expression is only an approximate scaling relation. The bottom expression corrects a missing factor of $1/\sqrt{2}$ in \citet[Eq. (9) therein]{bk2019}. 

\begin{figure}
    \centering
    \includegraphics{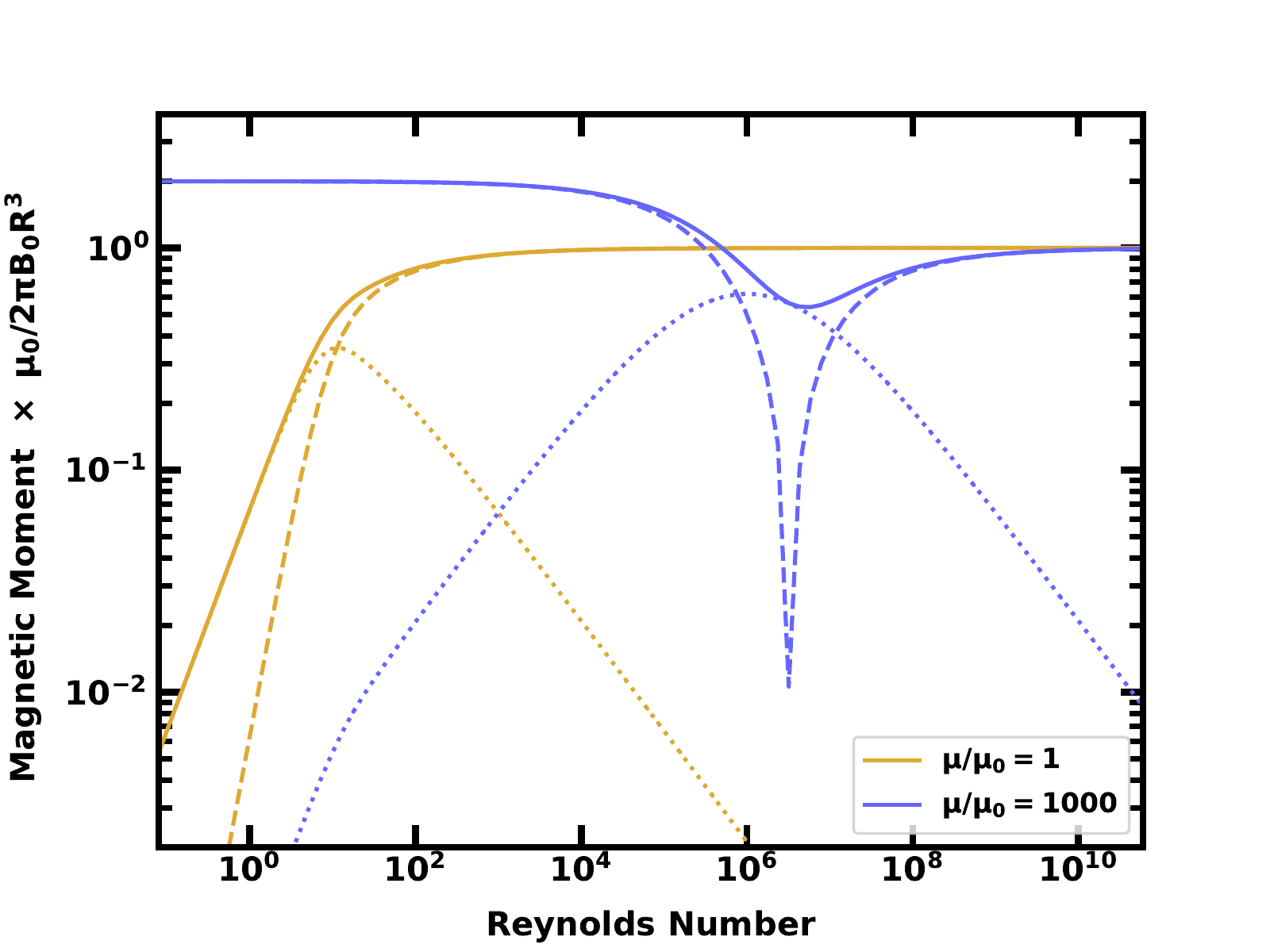}
    \caption{The magnetic moment of homogeneous conducting spheres in a sinusoidally varying magnetic field as a function of magnetic Reynolds number. The magnetic moment is scaled to a dimensionless quantity (e.g., Eq.~(\ref{eq:mmolim})). The gold curves are for a non-magnetic sphere, while the blue curves correspond to a sphere made of magnetic material. In each case, the line types distinguish the magnitude of the magnetic moment (solid curves) from the real (dashed) and imaginary (dotted) parts. Note that the real part is negative for the non-magnetic sphere, while for the magnetic sphere, it is positive at low Reynolds (below $R_m\sim \mu^2/\mu_0^2 \sim 10^6$) and negative at high Reynolds number. }
    \label{fig:spherforcereynoldsmmo}
\end{figure}

\begin{figure}
    \centering
    \includegraphics{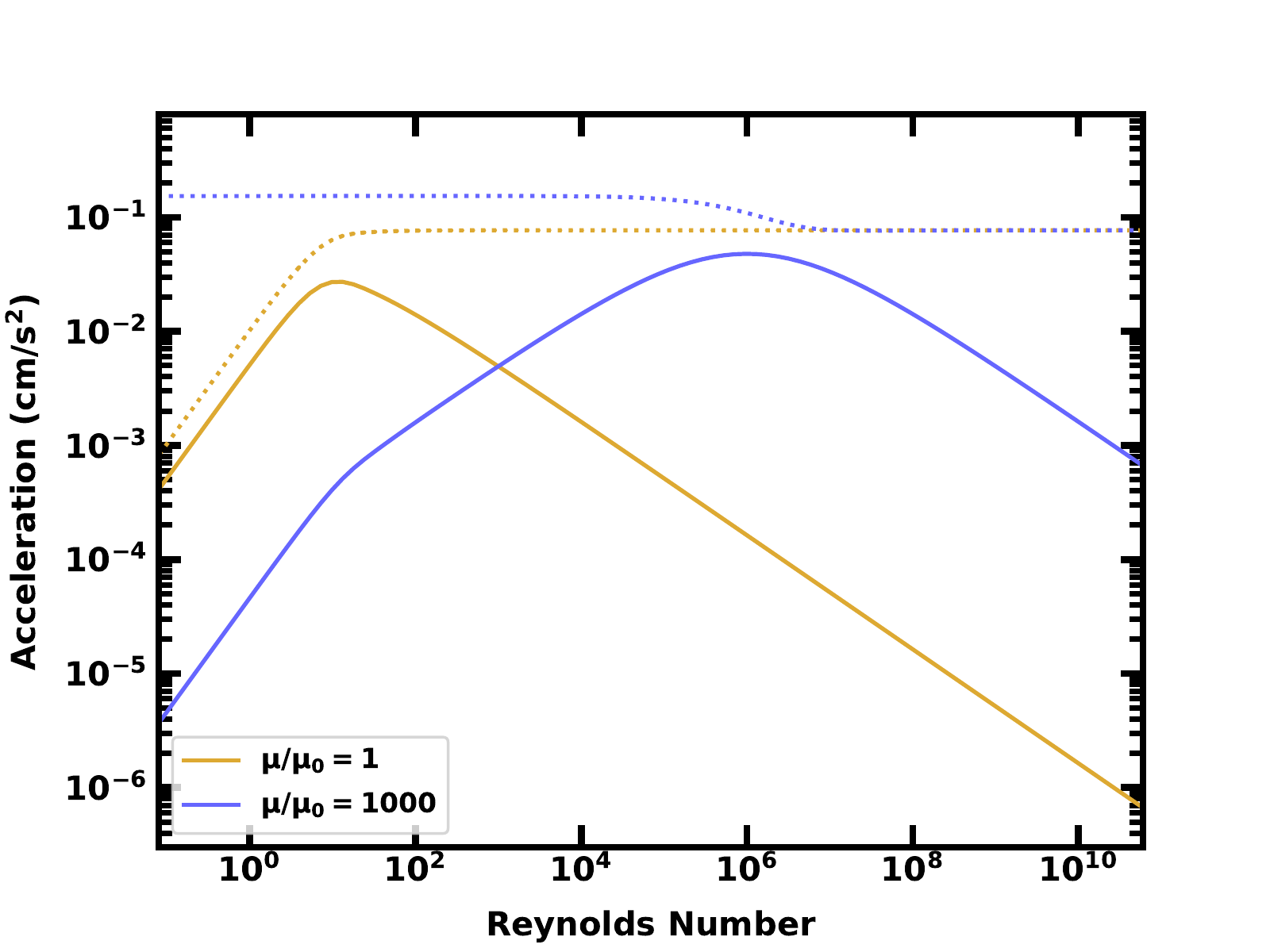}
    \caption{The acceleration of a conducting sphere traveling through a strong, spatially varying magnetic field as a function of magnetic Reynolds number. The field gradient exists only in the direction of travel. The Lorentz force is sinusoidal, with an average force that opposes the motion through the field, consistent with the power dissipated by Ohmic heating.  The gold curves show the acceleration of a non-magnetic sphere with constant conductivity; the solid curve is the average acceleration that opposed the direction of travel. The instantaneous acceleration has an oscillating part from the interplay between the induced magnetic moment of the conductor and the magnetic field. The dotted line shows the maximum value of the instantaneous acceleration. The blue curves correspond to a magnetic sphere with $\mu = 1000\mu_0$.}
    \label{fig:spherforcereynolds}
\end{figure}

Figure~\ref{fig:spherforcereynoldsmmo} provides an illustration of the magnetic moment of homogenous spherical conducting bodies as a function of Reynolds number. It shows examples of non-magnetic and magnetic bodies. The power dissipation in Ohmic heat is gleaned from the imaginary part of the each curve (Eq.~(\ref{eq:powahmmo})). Figure~\ref{fig:spherforcereynolds} gives examples of the instantaneous forces experienced by theses bodies.

\subsection{Comparison with motional EMF}

A conducting planet traveling through a stellar magnetic field experiences a charge separation from the Lorentz force independently of any induced currents. In a reference frame that moves with the planet, the stellar magnetic field lines flow past with some velocity $\vecvrel$, producing an electric field as a result of this motion:
\begin{equation}
    \vec{E} = \gamma \vecvrel\times \vec{B}, 
\end{equation}
where the Lorentz factor $\gamma$ may be ignored in the context of planetary orbital dynamics. Assuming an instantaneous response from charges within the conducting planet's interior, the planet is polarized with electric dipole moment
\begin{equation}
    \vec{p} = 4\pi \epsilon_0 R^3 \vec{E}.
\end{equation}
The force on the conductor then is approximately
\begin{eqnarray}
\vec{F}_E   & = &  \vec{p}\cdot\vec{\nabla}\vec{E}
\\ \label{eq:FE}
& = &  4\pi \epsilon_0 R^3 
\left(\vecvrel\times\vec{B}\right) \cdot \left(\vecvrel\times\vec{\nabla}\vec{B}  \right)
\\
\label{eq:FEx} 
& = & 4\pi \epsilon_0 R^3 \left[
\vrel^2 \, \vec{B} - (\vecvrel\cdot\vec{B}) \, \vecvrel
\right] \cdot \vec{\nabla}\vec{B}
\end{eqnarray}
To roughly compare this motional electromotive force (EMF) with the induction force, we use the ratio of magnitudes,
\begin{eqnarray}
    \frac{F_E}{F_B} & \sim &  \frac{\epsilon_0 R^3 \vrel^2 \kappa B^2}{ R^3 \kappa B^2/\mu_0} = \vrel^2\mu_0\epsilon_0 = \frac{\vrel^2}{c^2}  \ \ \ \ 
    (\mu \gg \mu_0 \ \ \text{or} \ \  
    R_m \gg \mu^2/\mu_0^2) \\
    & \sim &  \frac{\epsilon_0 R^3 \vrel^2 \kappa B^2}{ R^5 \cond \vrel \kappa^2 B^2} 
    \sim \frac{\vrel^2\epsilon_0\mu_0}{R^2 \mu_0 \cond v \kappa} 
    = \frac{\vrel^2}{c^2 R_m}
    \ \ \ \ (R_m \ll 10 \ \ \text{and} \ \ \mu \approx \mu_0),
\end{eqnarray}
 \myedit{where $\kappa$ is a wave number that characterizes the field variations along the planet's trajectory. For example, if the planet is on a circular, polar orbit with respect to a static stellar magnetic dipole, $\kappa \sim 1/R$. The product $\vrel \kappa$ thus represents the frequency $\omega$ of the field in the planet's frame of reference.} This comparison suggests that electric polarization is comparatively small when the magnetic Reynolds number is high, but that the polarization force can dominate over the magnetic force when $R_m$ is low, as with bodies that are small or made of non-magnetic, insulating material. We focus here only on scenarios where magnetic forces are more significant than those from motional EMF.

\section{Astrophysical examples}\label{sec:apex}

To apply the results of the previous sections, we consider three astrophysical examples: (i) bodies with fossil magnetic fields, (ii) those that are magnetizable, as in soft-iron spheres, and (iii) conducting asteroids and planets. Our goal is not to provide an exhaustive analysis, but to understand whether magnetic interactions plausibly lead to observable phenomena. We begin with fossil magnetic fields, focusing on compact, magnetized binary stars.

\subsection{A compact stellar binary with fossil fields}\label{sec:fossil}

Magnetic interactions between compact partners in a close binary pair may produce measurable dynamic effects during a merger event or even earlier \citep{bourgoin2022}. To assess the impact of magnetic interactions on orbits well before merging, we assume the partners have 'fossil' (permanent) magnetic dipole moments oriented antiparallel to each other and perpendicular to the orbital plane. Then, from the formalism in \S\ref{sec:magdyn}, the force experienced by one partner (labeled 'a') by the other star (labeled 'b') is  
\begin{eqnarray}
    \vec{F}_B & = & \vec{\mu}_a\cdot\vec{\nabla}\vec{B}_b \\
    \ & = & -\frac{12\pi B_{a} B_{b} R_a^3 R_b^3}{\mu_0 r^4}
    \left[ \left(1-5\frac{z^2}{r^2}\right)\uvec{r} + 2\frac{z}{r}\uvec{z}\right]\ \ \ \ (\vec{\mu}_a \parallel -\vec{\mu}_b)
\end{eqnarray}
where $r$ is the binary separation, and the positive $z$-axis is in the direction of $\vec{\mu}_b$, the magnetic moment of star 'b'. When $z=0$, the binary orbit is strictly in the plane perpendicular to the stellar magnetic moments. 

This force law leads to the possibility of an inspiral orbit. Following the analysis in \S\ref{subsec:magrmsco}, adapted for a pair of identical partners --- white dwarfs or neutron stars --- the magnetic minimum stable circular orbit is 
\begin{equation}\label{eq:rmsco}
r_\text{msco} = 2 \left[\frac{3 \pi}{\mu_0 G M_\star^2}\right]^{1/2} \frac{B_\star R_\star^3}{M_\star}
\end{equation}
With typical radii and masses of white dwarfs and neutron stars, along with surface field strengths near the maximum observed values $(R_\star,M_\star,B_\star) = (1\ \text{R}_\oplus, 1~\text{M}_\odot, 10^9$~G) and ($10\ \text{km}, 2~\text{M}_\odot, 10^{15}$~G) respectively, magnetic minimum stable circular orbits have values that are roughly 0.1\%\ of the stars' physical radii, formally placing these orbits deep within the stellar interiors. Even in these extreme astrophysical systems the field strengths are a few orders of magnitude too small to drive mergers.


At observed field strengths and orbital configurations, apsidal and nodal precession are potentially measurable.  When the magnetic dipoles are perpendicular to the orbital plane and antiparallel to each other, apsidal and nodal precession rates are
\begin{eqnarray}
    \dot{\varpi} = -\dot{\ascnode} & \approx & \frac{2^{1/2} 12\pi B_\star^2 R_\star^6}{\mu_0 G^{1/2} M_\star^{3/2} r^{7/2}},
\end{eqnarray}
which can yield precession rates for identical white dwarfs (WDs) or neutron stars in close binaries that are significant on dynamical time scales if the orbital separations are O(100) stellar radii or less. For example, with orbital and stellar parameters for the eclipsing WD-WD binary ZTF~J153932.16+502738.8 \citep{burdge2019}, 
\begin{equation}\label{eq:precess}
    \dot{\varpi} = -\dot{\ascnode} \approx 
    5.3 \fidval{B_a}{10^8\ \text{G}}\fidval{B_b}{10^8\ \text{G}} \ \text{deg/yr}
\end{equation}
Although the apsidal precession rate can be formally high, tidal forces and gravitational radiation circularize the binary orbit, preventing detection. Nodal precession, on the other hand, may occur if the fossil fields are not exactly antiparallel or are not strictly perpendicular to the  binary's orbital plane. The rate of precession is expected to be comparable the value in Equation~(\ref{eq:precess}). Light curves monitored over a period of time might provide novel constraints on the stellar magnetic fields.


\subsection{A magnetic body orbiting in the midplane of a fixed stellar dipole}

Next, we consider a magnetizable asteroid or planet ($\mu \gg \mu_0$) orbiting in the midplane of the stellar dipole. If either the sphere's radius or its conductivity are small, the Reynolds number is low and the magnetic field permeates the sphere. Then, the magnetic moment of the sphere is given by Equation~(\ref{eq:vecmmomagnetized}). Close-in orbits are of most interest, since the magnetic force falls off steeply with orbital distance ($r^{-7}$). The Roche radius gives an approximate lower limit to this distance; it is approximately
\begin{equation}\label{eq:roche}
R_\text{Roche} = K \left(\frac{M_\star}{\rho}\right)^{1/3} 
\approx 0.85 \fidval{K}{0.8}
\fidvalpow{M_\star}{1\ \Msolar}{1/3} \fidvalpow{\rho}{5\ \text{g/cm}^3}{-1/3}\ \Rsolar
\end{equation}
where $\rho$ is the orbiting body's mass density and $K$ is a constant near unity that depends on the material properties \citep[e.g.,][]{veras2017}.  Within this distance, tidal forces are destructive. 

Applying the stability analysis of \S\ref{subsec:magrmsco}, we equate the Roche radius to the magnetic minimum stable circular orbit radius $r_\text{msco}$ to find the stellar magnetic field strength needed to destabilize circular orbits. We start with the real part of the magnetic moment in Equation~(\ref{eq:mmolim}), applicable to a magnetizable, low-conductivity body. Then, we use $\xi_r = \mu r^{\gamma}$, where $\gamma = 3$ in this case, in  Equation~(\ref{eq:rmsco}). Finally, setting $r_\text{msco} = R_\text{Roche}$, we can solve for the required magnetic field:
\begin{eqnarray}
    B_\star & \gtrsim & \frac{R_\text{Roche}^{5/2}}{6 R_\star^3} \left[ \frac{\mu_0\rho M_\star (\mu+2\mu_0)}{ (\mu-\mu_0)}\right]^{1/2} \\
    & \approx &
 \frac{K^{5/2} M_\star^{4/3}}{6\rho^{1/3} R_\star^3} \left[ \frac{\mu_0 (\mu+2\mu_0)}{ (\mu-\mu_0)}\right]^{1/2}. 
\end{eqnarray}
For magnetic permeability significantly above unity, as for most ferromagnetic materials, the minimum destabilizing field is
\begin{equation}
B_\star
   \gtrsim 
    4.7 \times 10^{6} 
    \fidvalpow{K}{0.8}{5/2}
    \fidvalpow{\rho}{5\ \text{g/cm}^3}{-1/3}
    \fidvalpow{M_\star}{1~\Msolar}{4/3} 
    \fidvalpow{R_\star}{2~\Rsolar}{-3}
 \ \text{G}.
\end{equation}
This limit is two orders of magnitude higher than field strengths of $\sim$2~kG observed in T Tauri stars, to which the fiducial values of other parameters apply \citep[e.g.,][]{villebrun2019}. Similar assessments for white dwarfs indicate that their dipole fields are also too weak by orders of magnitude to destabilize circular orbits of magnetized material around them.  Increasing the material strength of the magnetic material, thereby reducing its Roche radius, lowers the requirement on magnetic field strength. Yet even a decrease in the Roche radius by a factor of 10 demands a field strength that is an order of magnitude higher than the most extreme values observed on white dwarfs. Neutron stars, even magnetars, have field strengths that are also too weak to destabilize these orbits. Finally, unstable orbits would have to be so close to the stellar host that it would be too hot to retain ferromagnetic properties.

Orbital precession, on the other hand, may occur at realistic field strengths. From \S\ref{subsec:orbsecular}, we find apsidal and nodal precession rates of 
\begin{align}
    \dot{\varpi} & \approx  -2.5 \dot{\ascnode} \approx \frac{45 B_\star^2 R_\star^6}{2 \rho G^{1/2}M_\star^{1/2}r^{13/2}}\frac{\mu-\mu_0}{\mu+2\mu_0} & \ 
    \\
    & \approx 3.0 
    \fidvalpow{\rho}{5\ \text{g/cm}^3}{-1}
    \fidvalpow{M_\star}{1~\Msolar}{1/2} 
    \fidvalpow{B_\star}{3~\text{kG}}{2}
    \fidvalpow{R_\star}{2~\Rsolar}{6}
    \fidvalpow{r}{0.01~\text{au}}{13/2}
    \ \text{deg/yr} &\text{(T Tauri)}
    \\
& \approx 4.4 
    \fidvalpow{\rho}{5\ \text{g/cm}^3}{-1}
    \fidvalpow{M_\star}{1~\Msolar}{1/2} 
    \fidvalpow{B_\star}{10^9~\text{G}}{2}
    \fidvalpow{R_\star}{1~\Rearth}{6}
    \fidvalpow{r}{0.2~\Rsolar}{13/2}
    \ \text{deg/yr} &\text{(WD)}
\end{align}
where the middle equation is for a body on a surface-skimming orbit around a T Tauri star and the bottom equation is for an orbit around a white dwarf. In both cases, $\mu \gg \mu_0$. The fiducial values for the white dwarf are extreme; the magnetic field strength is near the maximum observed value, and the orbital distance is well below the Roche radius in Equation~(\ref{eq:roche}), indicating that the magnet must be small ($\lesssim 100$~km) so that it is held together not by self-gravity but by its own material strength \citep[see][and references therein]{brouwers2022}. Under these conditions orbital precession is formally significant. These same conditions will also likely affect the orbiter's magnetic properties; high equilibrium temperatures near the stellar host will reduce the magnetic permeability.  While a magnetic asteroid may be able to maintain internal magnetization when cool, unlike larger bodies with internal heating \citep[see][for example]{kislyakova2020}, it is unlikely to remain solid or magnetic while occupying an orbit so close to the stellar host.

\subsection{Orbital dynamics at high Reynolds number: a conductor in a spinning stellar dipole}

We next consider the high Reynolds number limit, where induced currents and the Lorentz force play a role in the orbital dynamics of a conducting body. To explore this scenario, we assume that a rotating star has a magnetic dipole moment at right angles to its spin axis, so that the moment vector rotates in a plane. We further assume that a conducting planet is on a low-eccentricity orbit in that same plane.  When the planet is on a circular orbit, the planet's position as a function of time $t$ is
\begin{equation}\label{eq:r}
\vec{r} = \rcirc \left[\cos(\Omega t) \uvec{x} + \sin(\Omega t) \uvec{y}\right],
\end{equation}
where $\rcirc$ is the mean orbital distance of the planet. At that distance, the stellar magnetic moment evolves according to 
\begin{equation}
\vec{m}_\star = \frac{4\pi B_\star R_\star^3}{\mu_0}
\left[\cos(\Omega_\star t) \uvec{x} + \sin(\Omega_\star t) \uvec{y}\right];
\end{equation}
in these expressions, both the stellar magnetic moment and the planet's orbit lie in the $x-y$ plane. At the planet's location, the magnetic field is
\begin{eqnarray}
\vec{B}_\text{fix}(t) & = & B_\star \frac{R_\star^3}{\rcirc^3} 
\left\{ 3 \cos(\omega t) \left[\cos(\Omega t) \uvec{x} + \sin(\Omega t) \uvec{y}\right]  
- \left[\cos(\Omega_\star t) \uvec{x} + \sin(\Omega_\star t) \uvec{y}\right]
\right\}
\\
 & = & B_\star \frac{R_\star^3}{2 \rcirc^3} 
\left\{ 
\left[ 3 \cos(2\Omega t-\Omega_\star t) + \cos(\Omega_\star t) \right] \uvec{x} 
+ 
\left[ 3 \sin(2\Omega t-\Omega_\star t) + \sin(\Omega_\star t) \right] \uvec{y}  
\right\},
\end{eqnarray}
where $\omega \equiv \Omega_\star - \Omega$. 
We next measure the magnetic field at the planet's location when the planet is tidally locked to the star. We adopt a set of basis vectors that include $\uvec{r}$ (see Eq.~(\ref{eq:r})), a unit vector directed radially outward from the host star, along with the unit vector
\begin{equation}
    \uvec{v} = - \sin(\Omega t) \uvec{x} + \cos(\Omega t) \uvec{y}, 
\end{equation}
that is aligned with the planet's instantaneous direction of travel in the stellar rest frame. The local stellar magnetic field measured in this frame is 
\begin{eqnarray}\label{eq:Btidallylocked}
\vec{B}_\text{lock}(t) & = & (\vec{B}\cdot\uvec{r}) \uvec{r} + (\vec{B}\cdot\uvec{v}) \uvec{v} 
\\
\ & = &  B_\star \frac{R_\star^3}{\rcirc^3} 
\left[ 2 \cos(\omega t) \uvec{r} - \sin(\omega t) \uvec{v} \right].
\end{eqnarray}
In the planet's tidally-locked frame, there are again two modes of oscillation, although now with the same frequency. 

A similar analysis yields the Jacobian of the field in the orbital plane:
\begin{equation}\label{eq:Jacobianlockframe}
    \vec{\nabla}\vec{B}_\text{lock} = B_\star \frac{R_\star^3}{\rcirc^4}
    \left[-6\cos(\omega t) \uvec{r}\uvec{r} 
    + 2 \sin(\omega t) \uvec{r}\uvec{v} 
+ 3 \sin(\omega t) \uvec{v}\uvec{r} 
+ \cos(\omega t) \uvec{v}\uvec{v}
    \right]_\text{lock},
\end{equation}
where the subscript reminds that this expression applies in the tidally-locked frame.

Finally, the induced magnetic dipole moment in the high-Reynolds number limit (Eq.~\ref{eq:mmohicond}) is real-valued (in phase with the local stellar magnetic field), so that the acceleration experienced by the planet is
\begin{equation}
    \vec{a}_B =\frac{3 B_\star^2 R_\star^6}{4 \mu_0 \rho \rcirc^7  } 
\left\{ 
\left[9\cos(2\omega t) + 15\right] \uvec{r} - 3\sin(2 \omega t) \uvec{v}
\right\},
\end{equation}
\myedit{an expression which, we caution, is valid only when the planet experiences field variations rapid enough so that $R_m \gg 1$.}
As in Equation~(\ref{eq:powahmmo}), a purely real magnetic moment means negligible Ohmic losses. Thus, there is no net change of orbital energy. Furthermore, with a vanishing time-averaged acceleration, orbital precession from secular theory is insignificant.

\begin{figure}
    \centering
    \includegraphics{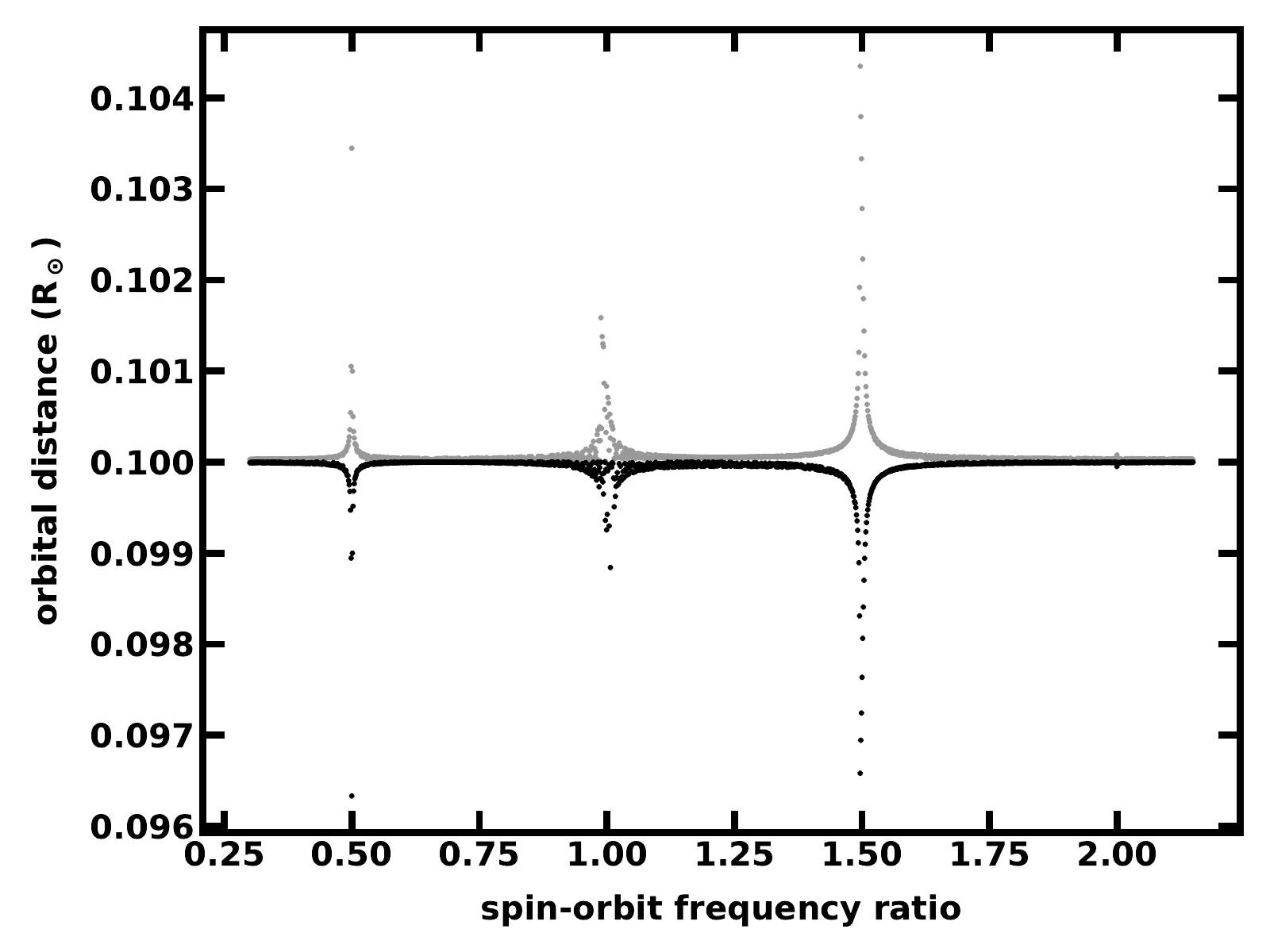}
    \caption{Simulated minimum and maximum radial excursion distances of a conducting asteroid as a function of the spin-orbit frequency ($\Omega_\star/\Omega$). The star is a 0.5~\Msolar, 1.5~\Rearth\ white dwarf with a $10^9$~G magnetic field that is oriented perpendicularly to the star's spin axis. The asteroid (1~km radius, density of 5~g/cm$^3$) is placed on an initially circular orbit at $0.1$~\Rsolar\ in the plane of the rotating stellar dipole moment. The asteroid's orbit acquires some small eccentricity over hundreds of orbits, as illustrated by the minimum and maximum radial excursions indicated by pairs of black and gray points, vertically displaced from one another. Each pair corresponds to a single orbit integration. At these orbital distances, the eccentricty ``pumping'' is weak, and occurs only near a spin-orbit resonance ( $\Omega_\star/\Omega \approx (1/2, 1, 3/2, 2$).}
    \label{fig:orbdyncirc}
\end{figure}

The oscillatory driving force, with a magnitude comparable to that experienced by a ferromagnet, has dynamical consequences nonetheless. To demonstrate, we adapted \orchestra, a hybrid $n$-body--coagulation code for planet formation \citep{bk2006, bk2011, kb2012, kb2014}, to include magnetic interactions. The new \orchestra\ code integrates the equations of motion for a conducting planet, initially on a circular orbit at two stellar radii from its magnetic host, over 250 orbits. In a suite of trials, the spin rate of the star is varied, and we estimate the planet's radial excursions \citep[e.g.,][]{sutherland2019} as a function of the spin rate of the star. Figure \ref{fig:orbdyncirc} illustrates the outcome in a plot of the minimum and maximum radial excursions versus the ratio of the star's spin frequency ($\Omega_\star$) to the orbital frequency ($\Omega$). We find that magnetic interactions can pump up the orbital eccentricity, depending on the ratio frequencies of stellar spin and orbital rotation. The radial excursions are broadest near resonances ($\Omega_\star:\Omega$ = 1:2, 1:1, 3:2, 2:1).

\begin{figure}
    \centering
    \includegraphics{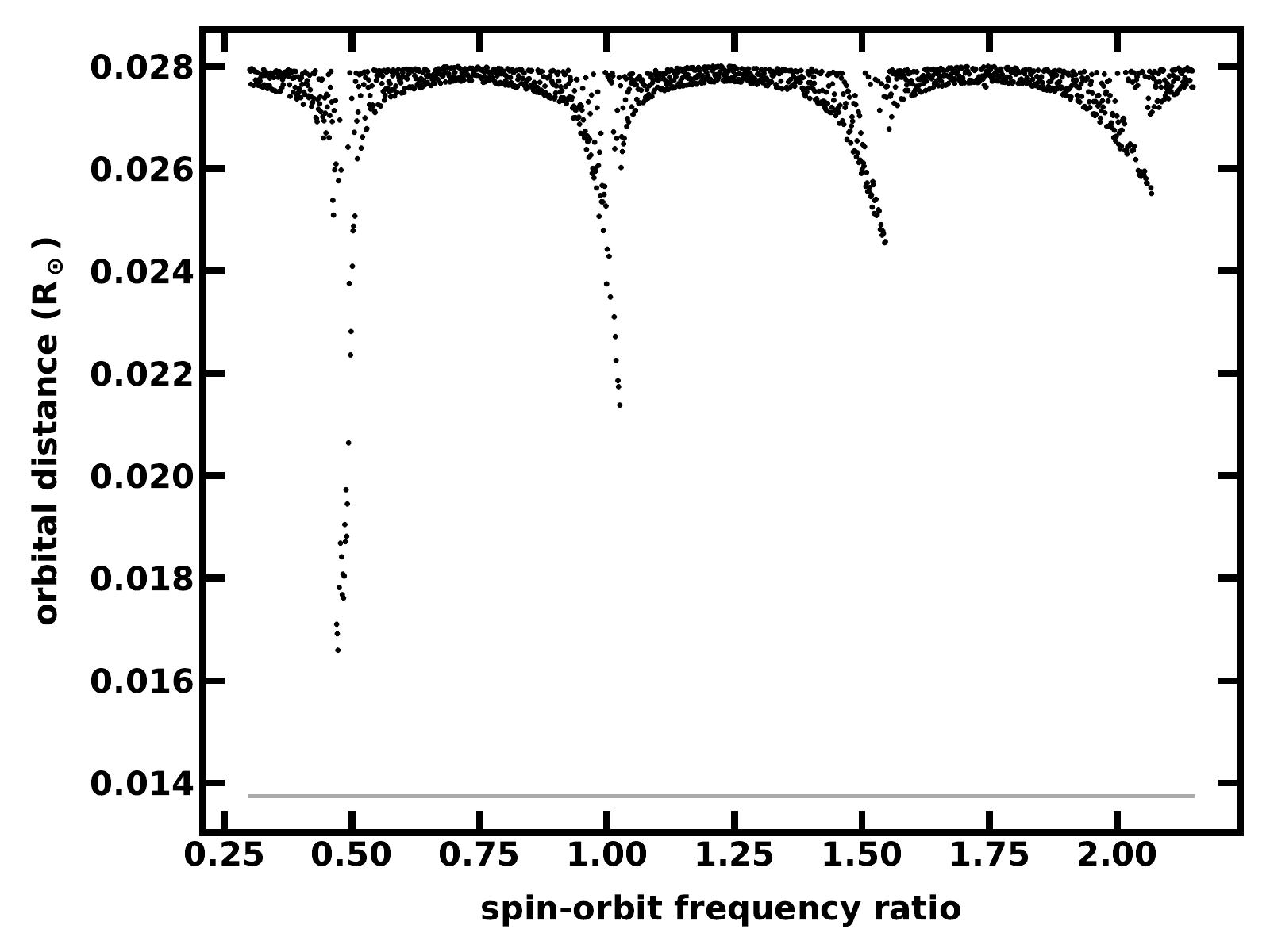}
    \caption{Simulated minimum radial excursion distances of a conducting asteroid as a function of the spin-orbit frequency ($\Omega_\star/\Omega$). The asteroid and host star have the same physical parameters as in the previous plot, Figure~\ref{fig:orbdyncirc}. Here, the asteroid is initially on a highly eccentric orbit with periastron at two stellar radii and apoastron at 1~\Rsolar. The eccentricity pumping near spin-orbit resonances is sufficient to significantly reduce the periastron distance, allowing a close approach to the stellar surface (the horizontal line at the bottom of the plot).}
    \label{fig:orbdynecc}
\end{figure}

Figure \ref{fig:orbdynecc} provides a second illustration of how magnetic interactions can impact orbits. It shows results from simulations of a conducting asteroid on a highly eccentric orbit about a white dwarf, with apoastron at 1~\Rsolar, and periastron at 3~\Rearth (two stellar radii). In some simulations, near spin-orbit resonances, the periastron distance evolves, bringing the asteroid close to the stellar surface. 

For an asteroid on an eccentric orbit, tidal dissipation could play an important role in orbital evolution. However, with peak losses only near closest approach to the star, the asteroid would tend to circularize near its original periastron distance over centuries, much longer than our simulation time. For example, a 1\%\ loss of orbital energy per orbit at 0.5~au would mean that the asteroid would have to absorb $O(10^{11})$~erg/g, which would vaporize it (assuming a specific heat capacity of $\sim 10^7$~erg/g/K). Our assumption here is that the asteroid is strong and rigid enough to survive each close passage. In contrast to this secular evolution, the resonant effects from the magnetic interaction can draw the asteroid closer to the stellar host on dynamical time scales. Still, other influences, including interactions with a gas disk and collisions, may compete with or overwhelm the magnetic phenomenon described here \citep{brouwers2022}.

Figures~\ref{fig:orbdyncirc} and \ref{fig:orbdynecc} illustrate effects that would be difficult to find in real astrophysical systems, even in the absence of other orbital perturbations. The host star must have a field strength that is near the peak of observed values and, even then, the conductor's orbit must be well within the formal Roche limit around the host. For a metallic object to survive at that location, we anticipate that the stellar host must be cool enough not to erode or evaporate the asteroid. Cool white dwarfs are known; an extreme example is DES J214756.46-403529.3 \citep{apps2021}. Its Gaia colors (G$_{BP}$-G$_{RP} \sim 2$) suggest a surface temperature above 3000~K. Yet, if the asteroid lingered at two stellar radii, it would certainly melt, losing the material strength it would need to withstand the tidal forces there. The asteroid could remain solid if it came close to the star only during periastron passage. Thus, the narrowest sliver of plausibility is allowed only by an unusually cold, highly magnetized white dwarf with a strong, conducting asteroid on a very eccentric orbit.

\begin{figure}
    \centering
    \includegraphics{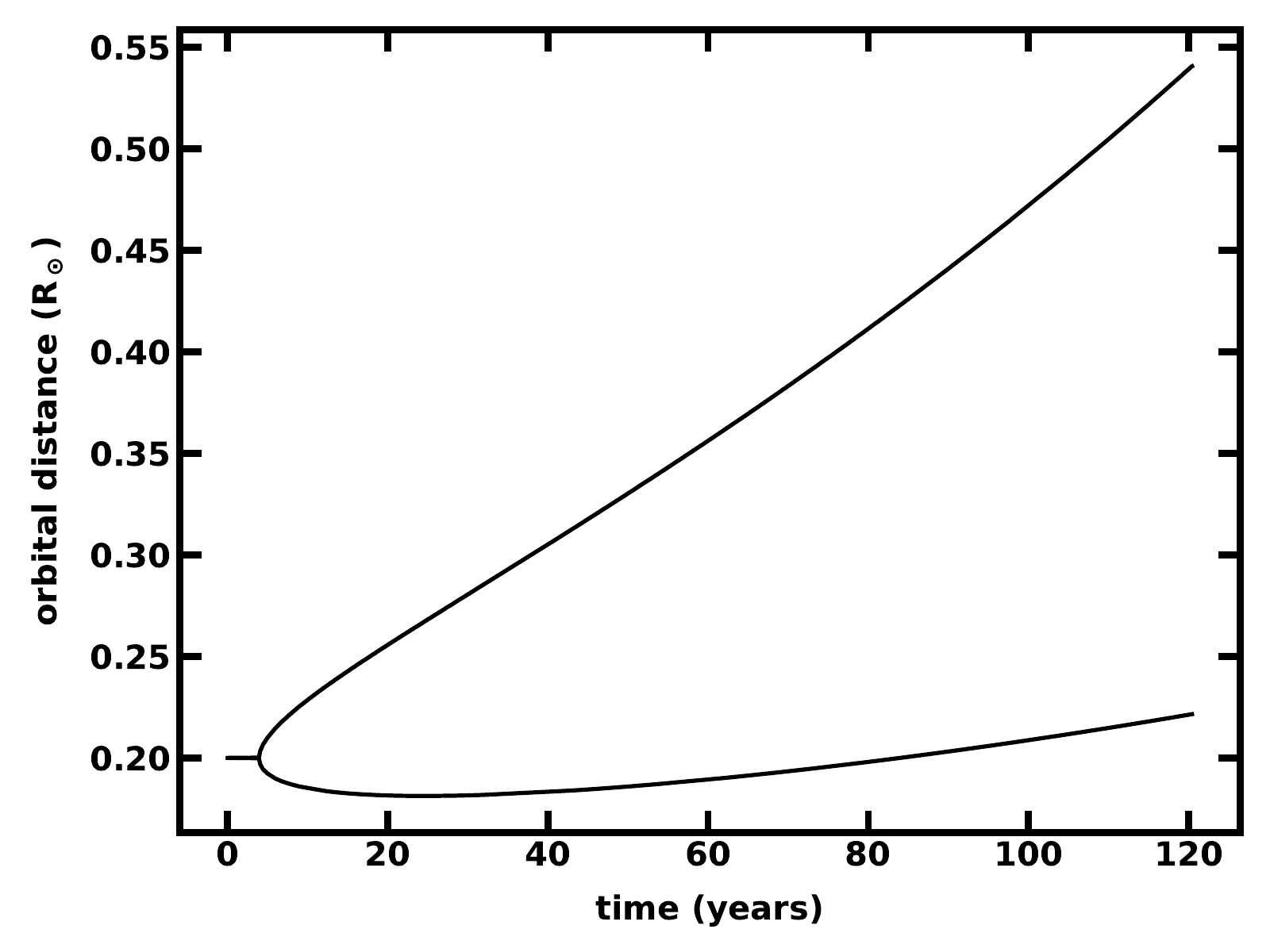}
    \caption{The radial excursions of a conducting asteroid trapped in a spin-orbit resonance. The white dwarf ($M_\star = 0.5~\Msolar$, $R_\star = 1.5~\Rearth$, and $B_\star = 10^9$~G) hosts an asteroid ($R = 1$~km, $\rho = 5$~g/cm$^3$) on an initially circular orbit at $0.2$~\Rsolar, as in Figure~\ref{fig:orbdyncirc}. This plot shows the evolution of the minimum and maximum radial excursions over a single orbit as the star's spin is slowly decreased. The spin-orbit frequency ratio is locked at 3:2 after the location of that resonance crosses the asteroid's initial path.}
    \label{fig:orbsweeprez}
\end{figure}

Spin-orbit interactions can drive orbital evolution if a conducting asteroid or planet becomes trapped in a resonance. For example, a young star with a strong magnetic field and rapid spin may experience spin-down, causing resonance locations to sweep outward. Conducting bodies in the sweep zone may get caught and pushed outward. In simulations, we find they experience eccentricity pumping as well. Figure~\ref{fig:orbsweeprez} provides an illustration of this phenomenon for an asteroid around a white dwarf. The spin-down rate is rapid --- the star's spin drops by a factor of almost two in a matter of a century. In astrophysical systems, inferred spin-down rates are much slower \citep[e.g.,][]{dejager1994, johnstone2015, johnstone2021}, with magnetars being the possible exception \citep[e.g.,][]{rea2010}. Our choice is for numerical convenience only; we observed no lower limit to the spin-down rate for resonant trapping.  On the contrary, trapping appears more robust for slower rates. Simulated conductors fell out of the resonant trap when spin-down was more rapid than in Figure~\ref{fig:orbsweeprez}.  

\subsection{Orbital evolution of a terrestrial planet in a spinning stellar dipole}

\begin{figure}
\centerline{\includegraphics{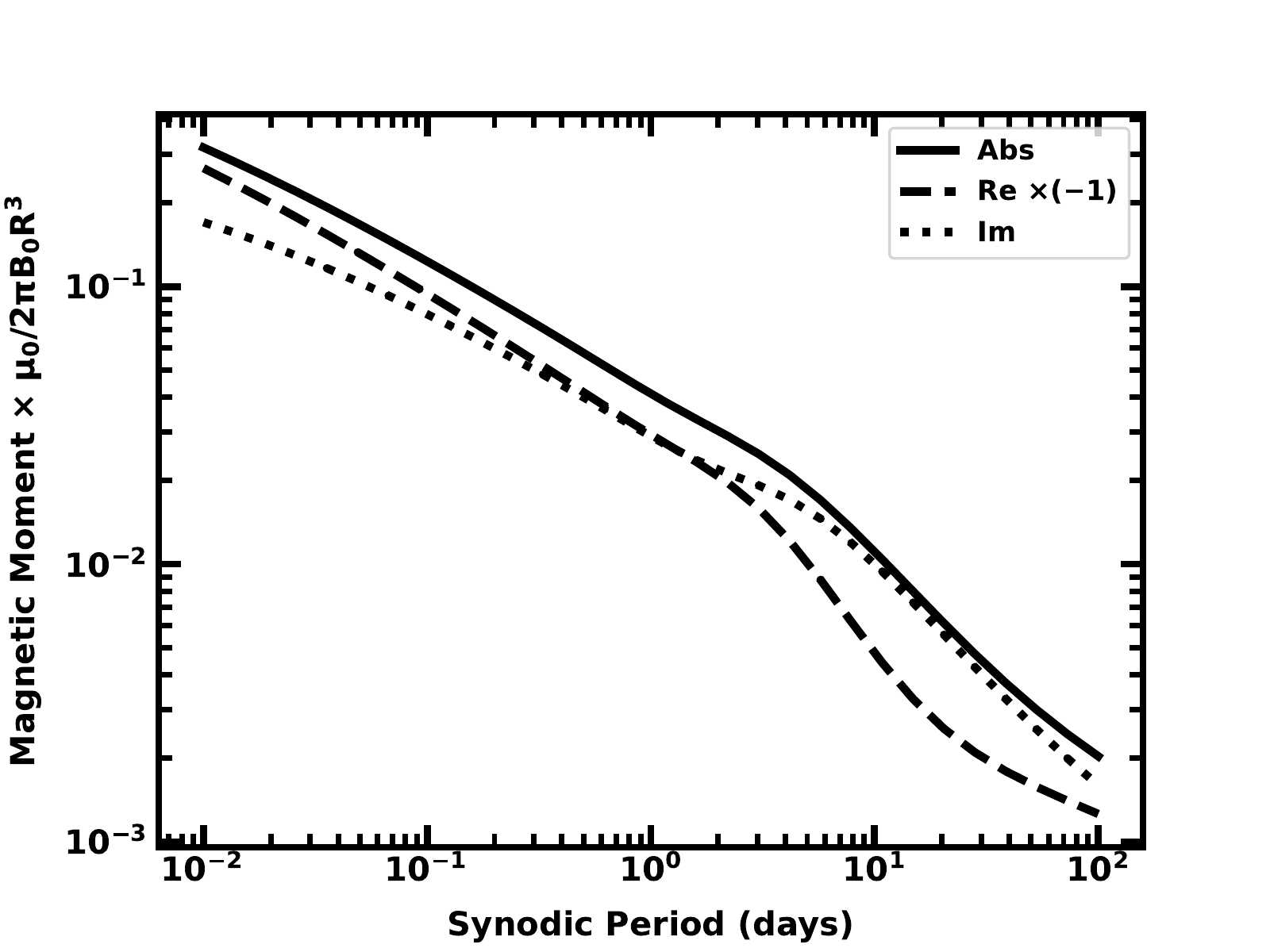}}
    \caption{\label{fig:merccond}The complex magnetic moment of a hypothetical Mercury, based on the parameters in Table~\ref{tab:mercstruct}, as a function of the period of oscillation of the stellar magnetic field at the planet's location. For each value of the period of oscillation, a grid-based calculation yields the complex magnetic moment vector. The grid consists of $10^5$ radial points; values of the conductivity on this grid are smoothed with an approximately Gaussian window with a standard deviation equivalent to 5~km (the thinnest radial zone is the crust, with a depth of 90~km). As in the legend, the solid line is the absolute magnitude of the planet's magnetic moment, the dashed line is the real part, times a factor of $-1$ since it is negative-valued (as an indicator of the phase shift between the applied and induced fields), and the dotted line is the imaginary part. Only the imaginary part leads to a net change in orbital energy as a result of Ohmic dissipation.
    }
\end{figure}

In this final example, we consider a more realistic composition for the orbiting body, modelled after the planet Mercury. Our hypothetical planet consists of a crust, mantle, inner core and outer core, each with unique conducting and magnetic properties as specified in Table~\ref{tab:mercstruct}. We assume spherical symmetry overall and apply the algorithm in Appendix~\ref{appx:powah} to determine the complex magnetic moment, $\vecmmo$. This calculation includes an assumed orbital distance and fundamental frequency of the oscillating field at the location of the planet, $\omega = \Omega_k-\Omega_\star$; Figure~\ref{fig:merccond} shows the magnetic moment for a range of frequencies. We also assume that the planet is tidally locked --- consistent with a close-in orbit where the magnetic field of the star is strong --- so that it has fixed orientation in the rotating reference frame of the star.\footnote{Planetary spin would affect the local magnetic flux calculations, and would be a source of energy in the Ohmic dissipation process. However, for the close-in orbits considered here, tidal locking is assumed.} Then, the planet's magnetic interactions involve two orthogonal magnetic field components, one directed radially outward from the star, and the other perpendicular to it in the orbital plane (Eq.~(\ref{eq:Btidallylocked})). These prescriptions allow us to completely specify the magnetic force on a ``Mercury'' in a close-in orbit. 

\begin{deluxetable}{lccccc}
\tabletypesize{\footnotesize}
\tablecolumns{12}
\tablewidth{0pt}
\tablecaption{Internal structure of Mercury.
    \label{tab:mercstruct}}
\tablehead{\colhead{\ } & \colhead{outer radius} & \colhead{thickness} & \colhead{density} & \colhead{conductivity} & \colhead{permeability}  
}
\decimals
\startdata
crust & 2440 km & 90 km & 2.8 g/cm$^3$ & 0.001 S/m & 1 $\mu_0$ \\
mantle & 2350 & 360 & 3.2 & 3.0 & 1 \\
outer core & 1990 & 1030 & 7.4 & $5.0\times 10^5$ & 1 \\
inner core & 960 & -- & 7.8 & $7.8 \times 10^5$ & 1000 
\enddata
\tablecomments{The radial distances and densities are representative of the results presented by \citet{genova2019}.  }
\end{deluxetable}


Writing the planetary magnetic moment as 
\begin{equation}
    \vecmmo = |\mmo| e^{-i\phi} \uvec{B},
\end{equation}
where  $\phi$ is a phase angle, and $\uvec{B}$ is a unit vector in the direction of the background magnetic field, we obtain the physical field 
\begin{equation}\label{eq:mmopolar}
    \vecmmop = |\mmo| \cos(\omega t + \phi) \uvec{B},
\end{equation}
where time $t$ is zero when the stellar magnetic dipole points at the planet's location. This equation carries all the information about how the magnetic dipole moment changes in time. Since the planet is tidally locked, there are two fields contributing, with different zero-points for time $t$, because the radial and tangential components of the stellar dipole peak at different locations.

The orbital evolution of this hypothetical Mercury broadly depends on how quickly it orbits the host star compared to the stellar rotation \citep{laine2008}. If the planet orbits more quickly than the star rotates, then it plows through the stellar magnetic field, experiencing instantaneous Lorentz forces including a net drag force that causes inspiral. Quantitatively, we can see this behavior by focusing on the imaginary part of the magnetic moment, which is positive valued, as it interacts with the radial component of the stellar magnetic field in the tidally locked frame of the planet. The time dependence of the dipole moment goes as $\sin\omega t$, which is 180 degrees out of phase with the gradient of the field in the direction of orbital motion. As in Equation~(\ref{eq:magforce}), the planet feels a force in opposition to its motion. 

When the stellar rotation is faster than the planet's orbit ($\omega < 0$), the planet perceives the stellar magnetic field lines as coming at it from the direction opposite to its orbital motion. Then, there is a net force in the direction of travel, boosting its orbital energy and pushing it outward. In quantitative terms, the sign of $\omega$ flips the sense of the sinusoidally varying magnetic moment, causing it to be in phase with the field gradient. This component of the magnetic force is thus directed along the planet's velocity vector.

When the planet corotates with the star, it experiences no change in the magnetic flux in its tidally locked frame; no currents are induced. The field is free to permeate through the interior of the planet, interacting only with the magnetic inner core. This configuration does not seem stable; if the planet drifts slightly inward, its new increased orbital speed will allow it to overtake the magnetic field lines, drawing it closer to the host star. If the planet is moved beyond the corotation radius, the stellar field lines overtake it, generating a torque that pushes out further still. 

A similar scenario plays out for binary stars with an important distinction. The Ohmic losses in a conductive companion as it interacts with the magnetic field of its rotating partner affect the partner's spin more than the orbital energy \citep{joss1979, campbell1983}. The torques from the magnetic interactions spin up the magnetic partner if the conducting star is within the corotation radius, pushing toward synchronization. Similarly, a conducting star beyond the corotation radius will spin down its magnetic partner, leading to synchronization. 

To establish how the magnetic force might impact a Mercury-like planet ($R = 2440$~km and $M = 3.30\times 10^{26}$~g) on a close-in orbit, we consider a stellar host with magnetic dipole moment vector lying in the plane perpendicular to the star's spin axis. The planet is on a circular orbit in that plane. The acceleration of the planet from interactions with the stellar magnetic field scales as
(Eq.~(\ref{eq:magforce}))
\begin{eqnarray}
    a_\text{mag} \sim \frac{\mmo}{M} B_\star\frac{R_\star^3}{\rcirc^4} 
    \sim \omega^{1.4} \rho B_\star^2 \frac{R_\star^6}{\rcirc^7}, 
\end{eqnarray}
where $\omega$ is the field variation frequency at the planet's location, and the power-law scaling is an approximation based on the data in Figure~\ref{fig:merccond}. 
%
Adopting $R_\star = 0.2~\Rsolar$, $M_\star = 0.2~\Msolar$, $B_\star = 8$~kG \citep[near the most extreme field strengths observed for M dwarfs, e.g.,][]{shulyak2017, shulyak2019, kochukhov2021}, an orbital distance of $\rcirc = 0.002$~au, and a frequency $\omega$ of the same order as the planet's orbital frequency ($2\pi$~d$^{-1}$), the planet is accelerated by roughly $10^{-8}$~cm/s$^2$ as a result of magnetic interactions. This value is roughly $\sim 10^{-12}$ times weaker than the magnitude of the gravitational acceleration.
Any orbital evolution as a result of magnetic interactions will be slow compared with the dynamical time.

Despite this comparative weakness, the magnetic interaction yields a continuous drag force on the hypothetical Mercury in a frame that corotates with the field lines. Over time, this force changes orbital energy; 
in terms of the orbital distance (Eq.~(\ref{eq:dadt})),
\begin{equation}
    \frac{d\rcirc}{dt} \sim 2 \frac{\Im\{\mmo\} B \Omega \rcirc^2}{GM_\star}
    \sim \Omega^{1.4} \rho B_\star^2 \frac{R_\star^6}{\rcirc^6}.
\end{equation}
The imaginary part of the magnetic moment associated with the average drag force is comparable to the magnitude of the magnetic moment itself in the case of the Mercury analog. In the example of a Sun-like star with a 10~kG magnetic field, the decay rate is 0.3~au/Gyr, suggesting inspiral  within tens of millions of years from a distance of a few Solar radii around a slowly rotating star. Table~\ref{tab:mercdyn} provides estimates of the orbital evolution time scales for a Mercury-like planet around a variety of stellar hosts. There, we use extreme values for the magnetic field strengths to illustrate what is possible, not what is typical.


\begin{deluxetable}{lcccccc}
\tabletypesize{\footnotesize}
\tablecolumns{12}
\tablewidth{0pt}
\tablecaption{Mercury around a variety of stellar hosts.
    \label{tab:mercdyn}}
\tablehead{\colhead{star type} & \colhead{$M_\star$} & \colhead{$R_\star$} & \colhead{$B_\star$} & \colhead{$P_\star$} & \colhead{$\rcirc$} & \colhead{$\rcirc/\dot{\rcirc}$}  
}
\decimals
\startdata
T Tauri & 1 M$_\odot$ & 2 R$_\odot$ & $3$ kG & $4$ day & 0.01 AU & -150 Myr \\
M dwarf & 0.2 M$_\odot$ & 0.2 R$_\odot$ & $8$ kG & $1$ day & 0.002 AU & -270 Myr \\
Ap/Bp & 3 M$_\odot$ & 3 R$_\odot$ & $30$ kG & $10$ day & 0.015 AU & -3 Myr \\
white dwarf & 0.5 M$_\odot$ & 1 R$_\oplus$ & $10^{9}$ G & $1$ day & 0.003 AU & -41 Myr \\
neutron star & 1.5 M$_\odot$ & 15 km & $10^{15}$ G & $10$ s & 0.004 AU & $+$3.9 Gyr \\
\enddata
\tablecomments{Stellar parameters are intended to be typical, except for the magnetic field strengths, which are characteristic of one the most extreme observed values for each type of star. For
T~Tauri stars see \citet{johnstone2014}; for M~dwarfs \citet{saar1985, johns-krull1996}, \citet{johnstone2014}, and \citet{kochukhov2021}. \citet{babcock1960} and \citet{landstreet1992} discuss Ap/Bp stars, and \citep{ferrario2020} summarizes observations of white dwarfs. \citep{olausen2014} introduce a catalog of neutron stars and magnetars, while \citet{younes2017} focuses on an extreme example. The orbital distance of the Mercury-like planet is nearest the larger of the stellar radius or the Roche radius (Eq.~(\ref{eq:roche})). The sign of the dynamical time indicates a growing ($+$) or shrinking ('-') planetary orbit.}
\end{deluxetable}

As in Table~\ref{tab:mercdyn}, orbital evolution time scales (last column) are most rapid for a Mercury-like planet around a peculiar A (Ap) star or white dwarf.
However, equilibrium temperatures at the close-in distances required for megayear dynamical times may well affect the assumed planetary structure for these calculations. Dynamical times around T Tauri stars, even at close-in distances, exceed the time that young stars spend in the T Tauri phase; as evolution proceeds, the magnetic field strength shrinks along with the stellar radius making long-term orbital evolution through magnetic interactions unlikely. Around a neutron star, a Mercury anaolog will be slowly pushed away from the Roche radius on gigayear time scales. Perhaps the most promising astronomical scenario is orbital evolution around a cool dwarf, at least early on in its evolution when it is active. Equilibrium temperatures are comparatively low, even for close-in orbits, and surface field strengths can exceed a kilogauss. Then, a close-in planet will be drawn inward to tidal destruction and accretion by the host star within a gigayear.

\section{Summary}

This work is an exploration of magnetic interactions between a host star and an orbiting companion. Previous work \citep[e.g.,][]{laine2008} demonstrated that induced currents and associated Ohmic heating losses cause secular changes in orbits, leading to inspiral in extreme cases. Here, we focus on instantaneous Lorentz forces that arise from interactions between a magnetic or conducting body and the stellar magnetic field. The following list summarizes our main findings:

\begin{itemize}
    \item The magnetic force between a star and an orbiting companion generally falls off more steeply than gravity, leading to the possibility of an unstable zone where mergers are inevitable. In astrophysical systems, the stellar magnetic fields have dipole field strengths that are orders of magnitude too weak to produce this type of unstable zone. At lower, more realistic magnetic field strengths, binaries with white dwarfs and/or neutron stars with fossil magnetic fields may cause measurable orbital precession \citep{bourgoin2022}.

    \item Conducting bodies in a time-varying stellar magnetic field develop eddy currents to suppress changes in magnetic flux. In the dipole-dipole interaction picture presented here, the magnetic moment of the body, $\vecmmo$, has a strength and phase that depend on the magnetic Reynolds number. At low $R_m$, the magnetic moment is 90$^\circ$ out of phase with the local stellar magnetic field; the work done on an orbiting body moving through a spatially varying magnetic field in this situation always opposes the motion \citep[e.g.,][]{giffin2010}. The amount of work rises with increasing $R_m$, as when the conductivity is higher, peaking at $R_m \sim 1$.  As $R_m$ increases beyond unity, the phase shifts toward 180$^\circ$, and the force becomes oscillatory, with little work done. The amplitude of the instantaneous force is strongest at high Reynolds number.
    
    \item The dipole-dipole interaction picture provides an estimate of the work done on a conducting body moving through a magnetic field that is consistent with calculations based on Ohmic dissipation \citep{joss1979, campbell1983, laine2008, laine2012, kislyakova2017, kislyakova2018, bk2019}. The dipole-dipole interaction view also allows for consideration of non-dissipative forces that also may impact orbital dynamics.

    \item When the magnetic Reynolds number of a conducting body is much greater than unity, the Ohmic losses are low. Then, the induced currents are in phase with the magnetic field itself; Lorentz forces are oscillatory. With these driving forces at play, orbital precession, eccentricity pumping, and resonant trapping are possible in specific configurations.

    \item Orbiting bodies with high magnetic susceptibility (as when $\mu/\mu_0 \gg 1$) and a magnetic Reynolds number that is much less than unity experience similar phenomena caused by oscillatory Lorentz forces.
    
    \item The most promising astrophysical systems for the observation of dipolar magnetic interactions are compact magnetic binaries. Fossil fields between close-in white dwarfs can cause significant orbital precession, a phenomenon also identified recently by \citet{bourgoin2022}. Nodal precession of misaligned fields may detectable even in circularized systems like ZTF~J153932.16+502738.8 \citep{burdge2019}, depending on the orientation of the stellar dipole moments.
\end{itemize}

These main conclusions are drawn on the basis of several simplifying assumptions. One is that the stellar magnetic fields a pure dipoles, falling off as $1/r^3$. This choice may underestimate the strength of the magnetic field around stars with winds \citep[e.g.,][]{johnstone2012}. Stellar winds can support and strengthen the field at larger radii, yielding a substantially shallower falloff, $1/r^2$. Depending on the azimuthal dependence of the wind-swept field, the effects described here may be stronger at large radii  Other plasma effects, including the formation of flux tubes \citep[e.g.,][]{goldreich1969, lai2012}, may modify the local magnetic field. These phenomena offer a potentially rich layer to the orbital dynamics that is well beyond what we describe here.

\acknowledgements We are grateful to referees for providing thoughtful comments that improved the content and presentation of this work. We acknowledge generous allotments of computer time on the NASA ‘discover’ cluster, provided by the NASA High-End Computing (HEC) Program through the NASA Center for Climate Simulation (NCCS). Guidance and comments from M.~Geller improved our presentation.


\software{Scipy \citep{scipy2001}}

\appendix 
\section{Induced currents in conducting asteroids or planets}\label{appx:powah}

In this section, we consider the currents induced in a conducting sphere by an oscillating magnetic field. The sphere has a radius $R$, is spherically symmetric with bulk properties, including mass density $\rho$, conductivity $\cond$, and magnetic permeability $\mu$, that may vary with radius. The external magnetic field in which the sphere resides is spatially uniform, with an amplitude that varies in time $t$ with frequency $\omega$, so that $\vec{B}_0 = B_0\exp(-i\omega t)\uvec{z}$, where $B_0$ is a constant. 

The changes in the magnetic field will induce currents in the conductor (Faraday's law), which will in turn generate a magnetic field that opposes the changes (Lenz's law). When the frequency $\omega$ is low (as quantified below), the induced currents are weak; the external field penetrates throughout the conducting body. When the magnetic field oscillations are rapid, the induced currents are strong and can generate an induced field that effectively prevents the external field from penetrating into the bulk of the conductor --- the skin effect. The key parameter in distinguishing these regimes is the magnetic Reynolds number, $R_m  \equiv  \mu\cond\omega R^2$ (Eq.~(\ref{eq:Rm})). For $R_m \ll 1$, the field varies slowly, and/or the conductivity and permeability are low; the external field bathes the entire conductor. When $R_m\gg 1$, the external field can only penetrate into a thin layer on the conductor's surface.

While we consider a range of frequencies, we assume throughout that the wavelength of electromagnetic waves within the sphere at frequency $\omega$ is much larger than the depth of field penetration into the sphere. This limitation on high frequencies allows us to ignore dielectric properties of the conductor \citep{bidinosti2007}. We begin with the case of a homogeneous, nearly-perfect conductor.

\subsection{A weakly conductive, non-magnetic sphere}

In the limit of low magnetic Reynolds number ($R_m\ll 1$), the external magnetic field is largely unaffected by the weak currents that it induces in the conducting body. We can use the `Physics II' formalism for Amp\`{e}re's law to estimate the current density:
\begin{equation}
    \int_r J(r,\theta)/\cond d\ell = i \pi \omega sin^2(\theta) r^2 B_0,   \ \ \ \ (r<R)
\end{equation}
where we have adopted spherical coordinates centered on the conductor and have assumed that the current density $J$ lies in the azimuthal direction.
Thus 
\begin{equation}
    J(r) =  \frac{i}{2} \cond\omega r \sin(\theta) B_0,
\end{equation}
where the factor of $i$ indicates that the current density lags the magnetic field by a phase of 90$^\circ$. The magnetic moment of the sphere is thus
\begin{equation}
    \vecmmo = \frac{1}{2}\int_V dV \vec{r}\times \vec{J}  = \frac{2\pi i}{15} \cond\omega R^5 \vec{B}_0.
\end{equation}
The power dissipated by the induced currents through Ohmic heating is 
\begin{equation}
    P_\Omega = \int_V dV \frac{\left|\vec{J}\right|^2}{\cond} 
    = \frac{2\pi}{15} \cond \omega^2 R^5 B_0^2.
\end{equation}
By associating the frequency $\omega$ with an orbital speed through a spatially varying magnetic field, we can deduce that Ohmic heating comes at the expense of orbital energy \citep{giffin2010, kislyakova2017, bk2019}.

\subsection{A highly conducting sphere}

In the limit of high conductivity, the penetration depth of the magnetic field into a non-magnetic spherical body is small. We follow \citet{joss1979}, focusing on a small patch on the sphere's surface with an infinitesimal planar surface element of area $dA$ that interacts with a uniform, applied magnetic field exterior to it. In the region beneath the conducting surface, Faraday's law and Amp\`ere's law are, respectively, 
\begin{equation}
    \vec{\nabla}\times\vec{J} = -i\omega\cond\vec{B}, \ \ \ \vec{\nabla}\times\vec{B} = \mu \vec{J},
\end{equation}
where $\vec{J}$ is the current density, and $\vec{B}$ is the magnetic field, including both the applied and induced components; here, $\omega$ is the driving frequency, $\cond$ is the conductivity and $\mu$ is the permeability inside the conductor, all of which we assume to be constant. We also have used the microscopic version of Ohm's law, ($\vec{J} = \cond \vec{E}$), to write the electric field $\vec{E}$ in terms of the free current density $\vec{J}$. The combination of the two equations gives
\begin{equation}\label{eq:faradayamperecomplex}
    \vec{\nabla}\times\left(\vec{\nabla}\times\vec{J}\right) = i\mu\cond\omega\vec{J}.
\end{equation}
Recognizing the (local) planar symmetry of the problem, we overlay a Cartesian grid $(x,y,z)$ on the surface of the conductor with $z>0$ specifying a vertical distance away from the conductor's surface. If we assume the magnetic field is parallel to the local surface, oriented along the $x$-axis, or simply just consider this component of the field, a solution emerges: 
\begin{equation}
    \vec{J} = J_0 \exp\left(-k|z|\right)\vec{e}_y
\end{equation}
where 
\begin{equation}
    k = \sqrt{i\mu\cond\omega} = (1+i)\sqrt{\mu\cond\omega/2}
\end{equation}
and $J_0$ is a constant. A boundary condition at the conducting surface is that the tangential field just beneath the surface, $B_\parallel^{(-)}/\mu$, equals $B_\parallel/\mu_0$, the tangential field above it. Then,
\begin{equation}\label{eq:Jhicond}
    J_0  = \frac{k B_{\parallel}^{+}}{\mu_0},
\end{equation}

The power dissipated per unit surface area is 
\begin{equation}
     \frac{dP}{dA}  = \int dz \frac{|J|^2}{\cond} = \frac{B_{\parallel}^2}{4\mu_0^2}\sqrt{\mu\cond\omega/2}.
\end{equation}

Zooming out to view the whole conductor as a sphere of radius $R$, we seek the tangential magnetic field at the conductor's surface, $B_{\parallel}$, by assuming that the surface currents create a uniform magnetization within the sphere, while outside of the sphere, these currents generate a magnetic dipole field:
\begin{eqnarray}
    \vec{B}(r<R) & = & \frac{2\mu_0}{4\pi R^3}\vec{m} + \vec{B}_0\\
    \vec{B}(r>R) & = & \frac{\mu_0}{4\pi R^3}\left[3(\vec{m}\cdot\hat{r})\uvec{r} - \vec{m}\right] + \vec{B}_0, 
\end{eqnarray}
where $R$ is the radius of the sphere, $\vecmmo$ is the induced dipole moment, and $B_0$ is the strength of the field far from the sphere.\footnote{A derivation from potential theory is straightforward; here we rely on uniqueness and that fact that the induced field inside the sphere must be uniform in order to cancel the ambient field. To match boundaries at the sphere's surface, the induced field outside of the sphere must be a dipole.} Since the field inside the conductor is zero, the magnetic moment is
\begin{equation}\label{eq:mmohicond}
    \vecmmo = -\frac{2\pi R^3}{\mu_0} \vec{B}_0.
\end{equation}
and the tangential component of the field just outside of the conductor's surface is
\begin{equation}
    B_{\parallel} = \frac{3}{2}\sin(\theta) B_0
\end{equation}
With this result, integration over the sphere's surface yields the total dissipated power,
\begin{equation}
    P_\Omega = \int d\phi \int \sin(\theta)d\theta \frac{dP}{dA} \sin^2(\theta)
   = \frac{3 \pi B_0^2}{\mu_0^2}\sqrt{\frac{\mu\omega}{2\cond}} R^2.
\end{equation}

This analysis is valid so long as the conductivity is high, and that the skin depth of magnetic diffusion into the sphere,
\begin{equation}
    \delta = \sqrt{2/\mu\cond\omega},
\end{equation}
is very small compared with $R$. 

\subsection{Homogeneous material with arbitrary conductivity and permeability}

A similar treatment to the one above allows us to consider the more general case of arbitrary conductivity in a homogeneous material where the magnetic field can permeate deep within the sphere. Following \citet{bidinosti2007}, the current density satisfies the Faraday-Amp\`ere-Ohm Equation (\ref{eq:faradayamperecomplex}),
\begin{equation}\label{eq:FAOuniform}
    {\nabla}^2\vec{J} + k^2\vec{J} = 0.
\end{equation}
where the $\nabla^2$ operator is the vector Laplacian, and we have made use of Gauss' law,
\begin{equation}
 \vec{\nabla}\cdot \vec{J} = 0
\end{equation}
in the absence of free charges. As in the previous example of an almost-perfectly conducting sphere, symmetry requires the current density $\vec{J}$ to be toroidal, with a dependence on polar angle $\theta$ that matches that of the background magnetic field tangent to the surface of the sphere (Eq.~(\ref{eq:Jhicond})). Thus, we seek a solution of the form
\begin{equation}\label{eq:J}
    \vec{J} = f(r) \sin\theta \uvec{\phi}
\end{equation}
in standard spherical coordinates ($r$,$\theta$,$\phi$) that are aligned with the background magnetic field. Then, the azimuthal component of Equation~(\ref{eq:FAOuniform}) gives
\begin{equation}\label{eq:faradayampereradial}
    \frac{\partial^2f}{\partial r^2} + \frac{2}{r}\frac{\partial f}{\partial r} + 
    \left(k^2 - \frac{2}{r^2}\right)f = 0.
\end{equation}
The solution is 
\begin{equation}\label{eq:f}
    f(r) = A j_1(kr),
\end{equation}
where $j_n$ is a spherical Bessel function of order $n$, and we have enforced the boundary condition that $f(r)$ must be regular at the origin. 

We next obtain the constant $A$ by matching boundary conditions across the spherical conductor's surface at radius $R$.  From Faraday's law, the magnetic field components inside the sphere are 
\begin{equation}
    B_r =  \frac{2  A\mu }{k^2 r} j_i(kr) \cos\theta, \ \ \ 
    B_\theta  = -\frac{A\mu }{k^2 r} \left[{j_1(kr)} + k r j_1^\prime(kr)\right] \sin\theta, 
\end{equation}
and outside of the sphere ($r>R$),
\begin{equation}
    B_r =  (B_0 + 2 \tilde{m}/r^3) \cos\theta, \ \ \ 
    B_\theta  = (-B_0 + \tilde{m}/r^3) \sin\theta, 
\end{equation}
where $\tilde{m}$ is a constant related to the induced field from the sphere, which we take to be a dipole \citep{bidinosti2007}. The boundary conditions at the sphere's surface are that the radial component of $\vec{B}$ is continuous, as is the tangential component of $\vec{H} = \mu \vec{B}$. These two conditions allow us to solve for the two unknowns, $A$ and $\tilde{m}$, where only the former is needed to specify the current density.

Solving for $A$, and invoking identities relating spherical Bessel functions and their derivatives, \citet[see also \citealt{nagel2018}]{bidinosti2007} find that
\begin{equation}
    A = \frac{9B_0}{2k}\frac{i\mu\cond\omega}{(\mu + 2\mu_0) j_0(kR)+ (\mu - \mu_0)j_2(kR))}.
\end{equation}
With Equations~(\ref{eq:J} and (\ref{eq:f}), this expression completes the solution for the current density $J$.

The magnetic moment associated with the induced current is 
\begin{equation}
    \mmo = \frac{2\pi R^3 B_0}{\mu_0}
    \frac{2(\mu-\mu_0)j_0(kR) + (2\mu+\mu_0)j_2(kR)}{(\mu+2\mu_0)j_0(kR) + (\mu-\mu_0)j_2(kR)}
\end{equation}
\citep{bidinosti2007}.

The average dissipated power comes from Ohm's Law in microscopic form,
\begin{equation}\label{eq:powah}
    P_\Omega =  \frac{1}{2} \int_{V} dV \frac{|\vec{J}|^2}{\cond}, 
\end{equation}
which we express  most generally in the form
\begin{equation}
    P_\Omega  \equiv   \frac{B_0^2}{2\mu_0} \omega \frac{4\pi R^3}{3} {\cal{F}} \\
\end{equation}
\citep{bk2019}. For the homogeneous, linear material considered in this section, 
\begin{eqnarray}
    {\cal{F}} = \Im\left\{\frac{9\mu [j_0(kR) + j_2(kR)]}{2[(\mu + 2\mu_0) j_0(kR) + (\mu-\mu_0) j_2(kR)]}\right\}.
\end{eqnarray}
The dissipated power's behavior falls into several regimes, depending on Reynolds number $R_m$ and the magnetic permeability, $\mu$.  In the first regime, the Reynolds number is low, with $R_m \ll 10$; the magnetic field permeates throughout the conductor and is largely unperturbed by the induced currents. In a second regime, $R_m \gg \mu^2/\mu_0^2 \geq 10$; then, surface currents are produced that expel the magnetic field --- there is little penetration of the magnetic field into the conductor. For magnetic materials, with $\mu>\mu_0$, there is an intermediate regime, $10 \lesssim R_m \lesssim \mu^2/\mu_0^2$, where the field permeates the conducting material, but bound currents throughout the medium also respond to the field. 

To get numerical values for quantities presented here, the Python \texttt{SciPy.special} module can evaluate Bessel functions with complex arguments. In some cases, the multiple-precision capabilities of the \texttt{mpmath} module are required. We recommend using multiple-precision arithmetic for all equations in this section. 

\subsection{Radially varying conductivity and permeability}

\citet{laine2008} and \citet[and references therein]{kislyakova2017} describe numerical approaches to calculate the power dissipated in a conducting sphere with radially varying conductivity in a time-varying magnetic field. Here, we follow a similar approach, while also accommodating radial variations in magnetization.  Our starting point is the Faraday-Amp\`{e}re-Ohm equation (\ref{eq:faradayampereradial}), now modified to accommodate variable $\mu$ and $\cond$:
\begin{equation}\label{eq:FAOfull}
\vec{\nabla} \times \left(\frac{1}{\mu} \vec{\nabla} \times \frac{\vec{J}}{\sigma}\right) = i\omega \vec{J}.
\end{equation}
A few choice vector identities applied to the $\phi$ component of the above expression leads to
\begin{equation}\label{eq:FAOrad}
    \frac{\partial^2f}{\partial r^2} + 
    \left(\frac{2}{r} + g_1\right) \frac{\partial f}{\partial r} + 
    \left(k^2 - \frac{2}{r^2} + \frac{g_1}{r} + g_2\right)f 
\end{equation}
where $|\vec{J}| = f \sin(\theta)$, as above, and 
\begin{align}
\label{eq:g1}
    g_1 &= -\frac{1}{\mu}\frac{\partial\mu}{\partial r} - \frac{2}{\cond}\frac{\partial\cond}{\partial r}
    \\ \label{eq:g2}
    g_2 &= \frac{1}{\mu\cond}\frac{\partial\mu}{\partial r}\frac{\partial\cond}{\partial r}
    + \frac{2}{\cond^2}\left(\frac{\partial\cond}{\partial r}\right)^2
    -\frac{1}{\sigma}\frac{\partial^2\cond}{\partial r^2}.
\end{align}
We solve Equation~(\ref{eq:FAOfull}) numerically by discretizing the radial domain: $\vec{r}\rightarrow \{r_j\}$ for $j = 0,1,2,\dots, N-1$, where $r_0 \sim \Delta r$, $r_j = j\Delta r + r_0$, and $r_{N-1} = R$ are points on an equally-spaced mesh. Converting the derivatives in Equation~(\ref{eq:FAOrad}) to finite differences, the expression takes on a matrix form,
\begin{equation}\label{eq:Afb}
{\mathbb A} \tilde{f} = \tilde{b}
\end{equation}
schematically, the matrix ${\mathbb A}$ and the two vectors $\tilde{f}$ and $\tilde{b}$ are
\begin{equation}
{\mathbb A} = \left[\begin{array}{ccccc}
1 & 0 & \dots & \ & 0 \\
a^-_1 & d_1 & a^+_1 & 0 & \dots \\
0 & a^-_2 & d_2 & a^+_2 & \dots \\
\vdots & \ & \ & \ddots & \ \\
0 & \dots & 0 & 0 & 1
\end{array}\right]\!,
\ \ \ \tilde{f} = \left[\begin{array}{c}
f_0 \\ f_i \\ \vdots \\ f_{N-2} \\ f_{N-1}
\end{array}\right]\!
\ \ \ 
\text{and}\ \ \ \tilde{b} = \left[\begin{array}{c}
f_\text{inner} \\ 0 \\ \vdots \\ 0 \\ f_\text{outer}
\end{array}\right];
\end{equation}
here, the subscripts correspond to radial positions, so that $f_j = f(r_j)$. The elements of the tridiagonal matrix are 
\begin{align}
\nonumber
a^\pm_j &= 1/\Delta r^2 \pm (2/r_j + g_{1,j})/2\Delta r \\
d_j &= -2/\Delta r^2 + k_j^2 - 2/r_j^2+ g_{1,j}/r_j + g_{2,j} 
\end{align}
where $g_{i,j}$ are from Equations~(\ref{eq:g1}) and (\ref{eq:g2}), derived either analytically from $\mu(r)$ and $\cond$, or with their numerical (finite-difference) derivatives.

The two remaining undefined constants in the matrix equation~(\ref{eq:Afb}) are the first and last elements of the RHS vector $\tilde{b}$, $b_0 = f_\text{inner}$ and $b_{N-1} = f_\text{outer}$, corresponding to Dirichlet boundary conditions. We set $f_\text{inner} = 0$, since the current density vanishes at the origin. (In practice, the innermost radial grid point is close to but not exactly at the origin, because of a coordinate singularity there.)

To obtain $f_\text{outer}$, we follow this simple plan: We first assume a trial value for the current density at the outer surface with $f^t(R) = 1$. We then solve the matrix equation
\begin{equation}\label{eq:fAinvb}
\tilde{f}^t = {\mathbb A}^{-1} \tilde{b}^t.
\end{equation}
where $b^t_{N-1} = 1$ and all other elements are zero. We derive the magnetic field at the surface such that key boundary conditions from Maxwell's equations are satisfied. The tridiagonal form of ${\mathbb A}$ allows for fast inversion compared with a general matrix; in our Python implementation, we use the SciPy \texttt{linalg} package and the \texttt{solve\_banded} routine. The final step is to rescale the current density so that the derived background field matches the actual field, $\vec{B}_0$. 
%
The (rescaled) solution we are after is
\begin{equation}
\tilde{f} = C \tilde{f}^t,
\end{equation}
where $C$ is a complex constant. To obtain $C$, we assume that the magnetic field has the form
\begin{equation}
B_r(R^+,0) = B_0 + 2 B_M \ \ \ \text{and} \ \ \ B_\theta(R^+,\pi/2) = -B_I + B_M
\end{equation}
where the left equation is the purely radial field at the point just above the sphere's pole at $z=+R$, and the the right equation corresponds to the poloidal field just beyond the sphere's equator, and $B_M$ is the strength of the induced dipole field from the sphere just outside its equator at radius $R^+$. From Faraday's law, the current density and field just inside the conductor's surface are related by
\begin{equation}
B_r(R^-,0) = \frac{2f(R^-)}{i\cond\omega R}\ \ \ \text{and} 
\ \ \ 
B_\theta(R^-,\pi/2) = -\frac{f(R^-) + R f^\prime(R^-)}{i\cond\omega R}.
\end{equation}
The boundary conditions that the radial component of $\vec{H} \equiv \vec{B}/\mu$ and the tangential component of $\vec{B}$ are continuous across the boundary yield two independent equations with two unknowns. Eliminating $B_M$, we find 
\begin{equation}
B = \frac{1}{3}\left(B_r(R^-,0) - 2 \frac{\mu_0}{\mu} B_\theta(R^-,\pi/2)\right) 
= \frac{2}{3}\frac{(\mu + \mu_0) f(R^-) + R \mu_0 f^\prime(R^-)} {i\mu\cond\omega R};
\end{equation}
since $f = C f^t$, we infer that 
\begin{equation}
C = \frac{3 i\mu\cond\omega R B_0}{2 (\mu + \mu_0) {f}(R^-) + 2 R \mu_0 {f}^\prime(R^-)}.
\end{equation}
In our numerical implementation, we choose $R^-$ to be $r_{N-2}$, and evaluate the above expression with a finite difference operation:
\begin{equation}
C = \frac{3 i\mu_{N-2}\cond_{N-2}\omega r_{N-2} B_0}{2 (\mu_{N-2} + \mu_0) f^t_{N-2} + 2 r_{N-2} \mu_0 (f^t_{N-1}-f^t_{N-3})/2\Delta r},
\end{equation}
which then multiplies all $f^t_j$ to yield the full set, $\tilde{f}$. This direct method is an alternative to iterative approaches of \citet[Newton-Raphson-Kantorovich]{laine2008} and \citet[recursive application of boundary conditions on radial shells]{kislyakova2017}.

\bibliography{msrev2}{}

\begin{thebibliography}{}
\expandafter\ifx\csname natexlab\endcsname\relax\def\natexlab#1{#1}\fi
\providecommand{\url}[1]{\href{#1}{#1}}
\providecommand{\dodoi}[1]{doi:~\href{http://doi.org/#1}{\nolinkurl{#1}}}
\providecommand{\doeprint}[1]{\href{http://ascl.net/#1}{\nolinkurl{http://ascl.net/#1}}}
\providecommand{\doarXiv}[1]{\href{https://arxiv.org/abs/#1}{\nolinkurl{https://arxiv.org/abs/#1}}}

\bibitem[{{Angel}(1978)}]{angel1978}
{Angel}, J.~R.~P. 1978, \araa, 16, 487,
  \dodoi{10.1146/annurev.aa.16.090178.002415}

\bibitem[{Apps {et~al.}(2021)Apps, Smart, \& Silvotti}]{apps2021}
Apps, K., Smart, R.~L., \& Silvotti, R. 2021, Research Notes of the {AAS}, 5,
  229, \dodoi{10.3847/2515-5172/ac2df2}

\bibitem[{{Babcock}(1958)}]{babcock1958}
{Babcock}, H.~W. 1958, \apjs, 3, 141, \dodoi{10.1086/190035}

\bibitem[{{Babcock}(1960)}]{babcock1960}
---. 1960, \apj, 132, 521, \dodoi{10.1086/146960}

\bibitem[{Bidinosti {et~al.}(2007)Bidinosti, Chapple, \&
  Hayden}]{bidinosti2007}
Bidinosti, C., Chapple, E., \& Hayden, M. 2007, Concepts in Magnetic Resonance
  Part B: Magnetic Resonance Engineering, 31B, 191,
  \dodoi{https://doi.org/10.1002/cmr.b.20090}

\bibitem[{{Bourgoin} {et~al.}(2022){Bourgoin}, {Le Poncin-Lafitte}, {Mathis},
  \& {Angonin}}]{bourgoin2022}
{Bourgoin}, A., {Le Poncin-Lafitte}, C., {Mathis}, S., \& {Angonin}, M.~C.
  2022, \prd, 105, 124042, \dodoi{10.1103/PhysRevD.105.124042}

\bibitem[{{Bromley} \& {Kenyon}(2006)}]{bk2006}
{Bromley}, B.~C., \& {Kenyon}, S.~J. 2006, \aj, 131, 2737,
  \dodoi{10.1086/503280}

\bibitem[{{Bromley} \& {Kenyon}(2011)}]{bk2011}
---. 2011, \apj, 731, 101, \dodoi{10.1088/0004-637X/731/2/101}

\bibitem[{{Bromley} \& {Kenyon}(2019)}]{bk2019}
---. 2019, \apj, 876, 17, \dodoi{10.3847/1538-4357/ab12e9}

\bibitem[{{Brouwers} {et~al.}(2022){Brouwers}, {Bonsor}, \&
  {Malamud}}]{brouwers2022}
{Brouwers}, M.~G., {Bonsor}, A., \& {Malamud}, U. 2022, \mnras, 509, 2404,
  \dodoi{10.1093/mnras/stab3009}

\bibitem[{{Burdge} {et~al.}(2019){Burdge}, {Coughlin}, {Fuller}, {Kupfer},
  {Bellm}, {Bildsten}, {Graham}, {Kaplan}, {Roestel}, {Dekany}, {Duev},
  {Feeney}, {Giomi}, {Helou}, {Kaye}, {Laher}, {Mahabal}, {Masci}, {Riddle},
  {Shupe}, {Soumagnac}, {Smith}, {Szkody}, {Walters}, {Kulkarni}, \&
  {Prince}}]{burdge2019}
{Burdge}, K.~B., {Coughlin}, M.~W., {Fuller}, J., {et~al.} 2019, \nat, 571,
  528, \dodoi{10.1038/s41586-019-1403-0}

\bibitem[{{Caiazzo} {et~al.}(2021){Caiazzo}, {Burdge}, {Fuller}, {Heyl},
  {Kulkarni}, {Prince}, {Richer}, {Schwab}, {Andreoni}, {Bellm}, {Drake},
  {Duev}, {Graham}, {Helou}, {Mahabal}, {Masci}, {Smith}, \&
  {Soumagnac}}]{caiazzo2021}
{Caiazzo}, I., {Burdge}, K.~B., {Fuller}, J., {et~al.} 2021, \nat, 595, 39,
  \dodoi{10.1038/s41586-021-03615-y}

\bibitem[{{Campbell}(1983)}]{campbell1983}
{Campbell}, C.~G. 1983, \mnras, 205, 1031, \dodoi{10.1093/mnras/205.4.1031}

\bibitem[{{Chang} {et~al.}(2012){Chang}, {Bodenheimer}, \& {Gu}}]{chang2012}
{Chang}, Y.-L., {Bodenheimer}, P.~H., \& {Gu}, P.-G. 2012, \apj, 757, 118,
  \dodoi{10.1088/0004-637X/757/2/118}

\bibitem[{{Christensen}(2010)}]{christensen2010}
{Christensen}, U.~R. 2010, in Heliophysics: Evolving Solar Activity and the
  Climates of Space and Earth, ed. C.~J. {Schrijver} \& G.~L. {Siscoe}
  (Cambridge: Cambridge University Press), 179--216

\bibitem[{{Chyba} \& {Hand}(2021)}]{chyba2021}
{Chyba}, C.~F., \& {Hand}, K.~P. 2021, \apjl, 922, L38,
  \dodoi{10.3847/2041-8213/ac399d}

\bibitem[{{de Jager} {et~al.}(1994){de Jager}, {Meintjes}, {O'Donoghue}, \&
  {Robinson}}]{dejager1994}
{de Jager}, O.~C., {Meintjes}, P.~J., {O'Donoghue}, D., \& {Robinson}, E.~L.
  1994, \mnras, 267, 577, \dodoi{10.1093/mnras/267.3.577}

\bibitem[{{Donati} \& {Landstreet}(2009)}]{donati2009}
{Donati}, J.~F., \& {Landstreet}, J.~D. 2009, \araa, 47, 333,
  \dodoi{10.1146/annurev-astro-082708-101833}

\bibitem[{{Farihi}(2016)}]{farihi2016}
{Farihi}, J. 2016, \nar, 71, 9, \dodoi{10.1016/j.newar.2016.03.001}

\bibitem[{{Ferrario} {et~al.}(2020){Ferrario}, {Wickramasinghe}, \&
  {Kawka}}]{ferrario2020}
{Ferrario}, L., {Wickramasinghe}, D., \& {Kawka}, A. 2020, Advances in Space
  Research, 66, 1025, \dodoi{10.1016/j.asr.2019.11.012}

\bibitem[{Genova {et~al.}(2019)Genova, Goossens, Mazarico, Lemoine, Neumann,
  Kuang, Sabaka, Hauck~II, Smith, Solomon, \& Zuber}]{genova2019}
Genova, A., Goossens, S., Mazarico, E., {et~al.} 2019, Geophysical Research
  Letters, 46, 3625, \dodoi{https://doi.org/10.1029/2018GL081135}

\bibitem[{{Giffin} {et~al.}(2010){Giffin}, {Shneider}, {Kalra}, {Ames}, \&
  {Miles}}]{giffin2010}
{Giffin}, A., {Shneider}, M., {Kalra}, C.~S., {Ames}, T.~L., \& {Miles}, R.~B.
  2010, arXiv e-prints, arXiv:1004.5412.
\newblock \doarXiv{1004.5412}

\bibitem[{{Goldreich} \& {Lynden-Bell}(1969)}]{goldreich1969}
{Goldreich}, P., \& {Lynden-Bell}, D. 1969, \apj, 156, 59,
  \dodoi{10.1086/149947}

\bibitem[{{Grankin}(2021)}]{grankin2021}
{Grankin}, K. 2021, Acta Astrophysica Taurica, 2, 9,
  \dodoi{10.31059/aat.vol2.iss1.pp9-20}

\bibitem[{{Hogg} {et~al.}(2021){Hogg}, {Cutter}, \& {Wynn}}]{hogg2021}
{Hogg}, M.~A., {Cutter}, R., \& {Wynn}, G.~A. 2021, \mnras, 500, 2986,
  \dodoi{10.1093/mnras/staa3316}

\bibitem[{{Johns-Krull} \& {Valenti}(1996)}]{johns-krull1996}
{Johns-Krull}, C.~M., \& {Valenti}, J.~A. 1996, \apjl, 459, L95,
  \dodoi{10.1086/309954}

\bibitem[{{Johnstone}(2012)}]{johnstone2012}
{Johnstone}, C.~P. 2012, PhD thesis, Saint Andrews University, UK

\bibitem[{{Johnstone} {et~al.}(2021){Johnstone}, {Bartel}, \&
  {G{\"u}del}}]{johnstone2021}
{Johnstone}, C.~P., {Bartel}, M., \& {G{\"u}del}, M. 2021, \aap, 649, A96,
  \dodoi{10.1051/0004-6361/202038407}

\bibitem[{{Johnstone} {et~al.}(2015){Johnstone}, {G{\"u}del}, {Brott}, \&
  {L{\"u}ftinger}}]{johnstone2015}
{Johnstone}, C.~P., {G{\"u}del}, M., {Brott}, I., \& {L{\"u}ftinger}, T. 2015,
  \aap, 577, A28, \dodoi{10.1051/0004-6361/201425301}

\bibitem[{{Johnstone} {et~al.}(2014){Johnstone}, {Jardine}, {Gregory},
  {Donati}, \& {Hussain}}]{johnstone2014}
{Johnstone}, C.~P., {Jardine}, M., {Gregory}, S.~G., {Donati}, J.~F., \&
  {Hussain}, G. 2014, \mnras, 437, 3202, \dodoi{10.1093/mnras/stt2107}

\bibitem[{Jones {et~al.}(2001--)Jones, Oliphant, Peterson,
  {et~al.}}]{scipy2001}
Jones, E., Oliphant, T., Peterson, P., {et~al.} 2001--, {SciPy}: Open source
  scientific tools for {Python}.
\newblock \url{http://www.scipy.org/}

\bibitem[{{Joss} {et~al.}(1979){Joss}, {Katz}, \& {Rappaport}}]{joss1979}
{Joss}, P.~C., {Katz}, J.~I., \& {Rappaport}, S. 1979, \apj, 230, 176,
  \dodoi{10.1086/157074}

\bibitem[{{Katz}(1989)}]{katz1989}
{Katz}, J.~I. 1989, \mnras, 239, 751, \dodoi{10.1093/mnras/239.3.751}

\bibitem[{{Katz}(2017)}]{katz2017}
---. 2017, \apj, 835, 150, \dodoi{10.3847/1538-4357/835/2/150}

\bibitem[{{Kenyon} \& {Bromley}(2012)}]{kb2012}
{Kenyon}, S.~J., \& {Bromley}, B.~C. 2012, \aj, 143, 63,
  \dodoi{10.1088/0004-6256/143/3/63}

\bibitem[{{Kenyon} \& {Bromley}(2014)}]{kb2014}
---. 2014, \apj, 780, 4, \dodoi{10.1088/0004-637X/780/1/4}

\bibitem[{{Kenyon} \& {Bromley}(2017)}]{kb2017wd}
---. 2017, \apj, 850, 50, \dodoi{10.3847/1538-4357/aa9570}

\bibitem[{{Kepler} {et~al.}(2016){Kepler}, {Pelisoli}, {Koester}, {Ourique},
  {Romero}, {Reindl}, {Kleinman}, {Eisenstein}, {Valois}, \&
  {Amaral}}]{kepler2016}
{Kepler}, S.~O., {Pelisoli}, I., {Koester}, D., {et~al.} 2016, \mnras, 455,
  3413, \dodoi{10.1093/mnras/stv2526}

\bibitem[{{Khurana} {et~al.}(1998){Khurana}, {Kivelson}, {Stevenson},
  {Schubert}, {Russell}, {Walker}, \& {Polanskey}}]{khurana1998}
{Khurana}, K.~K., {Kivelson}, M.~G., {Stevenson}, D.~J., {et~al.} 1998, \nat,
  395, 777, \dodoi{10.1038/27394}

\bibitem[{{Kislyakova} \& {Noack}(2020)}]{kislyakova2020}
{Kislyakova}, K., \& {Noack}, L. 2020, \aap, 636, L10,
  \dodoi{10.1051/0004-6361/202037924}

\bibitem[{Kislyakova {et~al.}(2018)Kislyakova, Fossati, Johnstone, Noack,
  Lüftinger, Zaitsev, \& Lammer}]{kislyakova2018}
Kislyakova, K.~G., Fossati, L., Johnstone, C.~P., {et~al.} 2018, The
  Astrophysical Journal, 858, 105, \dodoi{10.3847/1538-4357/aabae4}

\bibitem[{{Kislyakova} {et~al.}(2017){Kislyakova}, {Noack}, {Johnstone},
  {Zaitsev}, {Fossati}, {Lammer}, {Khodachenko}, {Odert}, \&
  {G{\"u}del}}]{kislyakova2017}
{Kislyakova}, K.~G., {Noack}, L., {Johnstone}, C.~P., {et~al.} 2017, Nature
  Astronomy, 1, 878, \dodoi{10.1038/s41550-017-0284-0}

\bibitem[{{Kivelson} {et~al.}(2000){Kivelson}, {Khurana}, {Russell}, {Volwerk},
  {Walker}, \& {Zimmer}}]{kivelson2000}
{Kivelson}, M.~G., {Khurana}, K.~K., {Russell}, C.~T., {et~al.} 2000, Science,
  289, 1340, \dodoi{10.1126/science.289.5483.1340}

\bibitem[{{Kochukhov}(2021)}]{kochukhov2021}
{Kochukhov}, O. 2021, \aapr, 29, 1, \dodoi{10.1007/s00159-020-00130-3}

\bibitem[{{Konar}(2017)}]{konar2017}
{Konar}, S. 2017, Journal of Astrophysics and Astronomy, 38, 47,
  \dodoi{10.1007/s12036-017-9467-4}

\bibitem[{{Lai}(2012)}]{lai2012}
{Lai}, D. 2012, \apjl, 757, L3, \dodoi{10.1088/2041-8205/757/1/L3}

\bibitem[{{Laine} \& {Lin}(2012)}]{laine2012}
{Laine}, R.~O., \& {Lin}, D. N.~C. 2012, \apj, 745, 2,
  \dodoi{10.1088/0004-637X/745/1/2}

\bibitem[{{Laine} {et~al.}(2008){Laine}, {Lin}, \& {Dong}}]{laine2008}
{Laine}, R.~O., {Lin}, D. N.~C., \& {Dong}, S. 2008, \apj, 685, 521,
  \dodoi{10.1086/589177}

\bibitem[{{Landstreet}(1992)}]{landstreet1992}
{Landstreet}, J.~D. 1992, \aapr, 4, 35, \dodoi{10.1007/BF00873569}

\bibitem[{{Lavail} {et~al.}(2017){Lavail}, {Kochukhov}, {Hussain}, {Alecian},
  {Herczeg}, \& {Johns-Krull}}]{lavail2017}
{Lavail}, A., {Kochukhov}, O., {Hussain}, G.~A.~J., {et~al.} 2017, \aap, 608,
  A77, \dodoi{10.1051/0004-6361/201731889}

\bibitem[{{Malhotra}(1996)}]{malhotra1996}
{Malhotra}, R. 1996, \aj, 111, 504, \dodoi{10.1086/117802}

\bibitem[{{Mik{\'o}czi}(2021)}]{balazs2021}
{Mik{\'o}czi}, B. 2021, arXiv e-prints, arXiv:2109.10722.
\newblock \doarXiv{2109.10722}

\bibitem[{{Moreno} {et~al.}(2015){Moreno}, {Pichardo}, \&
  {Schuster}}]{moreno2015}
{Moreno}, E., {Pichardo}, B., \& {Schuster}, W.~J. 2015, \mnras, 451, 705,
  \dodoi{10.1093/mnras/stv962}

\bibitem[{{Nagel}(2018)}]{nagel2018}
{Nagel}, J.~R. 2018, IEEE Antennas and Propagation Magazine, 60, 81,
  \dodoi{10.1109/MAP.2017.2774206}

\bibitem[{{Noack} {et~al.}(2021){Noack}, {Kislyakova}, {Johnstone},
  {G{\"u}del}, \& {Fossati}}]{noack2021}
{Noack}, L., {Kislyakova}, K.~G., {Johnstone}, C.~P., {G{\"u}del}, M., \&
  {Fossati}, L. 2021, \aap, 651, A103, \dodoi{10.1051/0004-6361/202040176}

\bibitem[{{Olausen} \& {Kaspi}(2014)}]{olausen2014}
{Olausen}, S.~A., \& {Kaspi}, V.~M. 2014, \apjs, 212, 6,
  \dodoi{10.1088/0067-0049/212/1/6}

\bibitem[{{Piro}(2012)}]{piro2012}
{Piro}, A.~L. 2012, \apj, 755, 80, \dodoi{10.1088/0004-637X/755/1/80}

\bibitem[{{Rea} {et~al.}(2010){Rea}, {Esposito}, {Turolla}, {Israel}, {Zane},
  {Stella}, {Mereghetti}, {Tiengo}, {G{\"o}tz}, {G{\"o}{\u{g}}{\"u}{\c{s}}}, \&
  {Kouveliotou}}]{rea2010}
{Rea}, N., {Esposito}, P., {Turolla}, R., {et~al.} 2010, Science, 330, 944,
  \dodoi{10.1126/science.1196088}

\bibitem[{{Saar} \& {Linsky}(1985)}]{saar1985}
{Saar}, S.~H., \& {Linsky}, J.~L. 1985, \apjl, 299, L47, \dodoi{10.1086/184578}

\bibitem[{{Shulyak} {et~al.}(2017){Shulyak}, {Reiners}, {Engeln}, {Malo},
  {Yadav}, {Morin}, \& {Kochukhov}}]{shulyak2017}
{Shulyak}, D., {Reiners}, A., {Engeln}, A., {et~al.} 2017, Nature Astronomy, 1,
  0184, \dodoi{10.1038/s41550-017-0184}

\bibitem[{{Shulyak} {et~al.}(2019){Shulyak}, {Reiners}, {Nagel}, {Tal-Or},
  {Caballero}, {Zechmeister}, {B{\'e}jar}, {Cort{\'e}s-Contreras}, {Martin},
  {Kaminski}, {Ribas}, {Quirrenbach}, {Amado}, {Anglada-Escud{\'e}}, {Bauer},
  {Dreizler}, {Guenther}, {Henning}, {Jeffers}, {K{\"u}rster}, {Lafarga},
  {Montes}, {Morales}, \& {Pedraz}}]{shulyak2019}
{Shulyak}, D., {Reiners}, A., {Nagel}, E., {et~al.} 2019, \aap, 626, A86,
  \dodoi{10.1051/0004-6361/201935315}

\bibitem[{{Strugarek} {et~al.}(2017){Strugarek}, {Bolmont}, {Mathis}, {Brun},
  {R{\'e}ville}, {Gallet}, \& {Charbonnel}}]{strugarek2017}
{Strugarek}, A., {Bolmont}, E., {Mathis}, S., {et~al.} 2017, \apjl, 847, L16,
  \dodoi{10.3847/2041-8213/aa8d70}

\bibitem[{{Sutherland} \& {Kratter}(2019)}]{sutherland2019}
{Sutherland}, A.~P., \& {Kratter}, K.~M. 2019, \mnras, 487, 3288,
  \dodoi{10.1093/mnras/stz1503}

\bibitem[{{Valyavin}(2015)}]{valyvin2015}
{Valyavin}, G. 2015, in Astronomical Society of the Pacific Conference Series,
  Vol. 494, Physics and Evolution of Magnetic and Related Stars, ed. Y.~Y.
  {Balega}, I.~I. {Romanyuk}, \& D.~O. {Kudryavtsev}, 107

\bibitem[{{Veras} {et~al.}(2017){Veras}, {Carter}, {Leinhardt}, \&
  {G{\"a}nsicke}}]{veras2017}
{Veras}, D., {Carter}, P.~J., {Leinhardt}, Z.~M., \& {G{\"a}nsicke}, B.~T.
  2017, \mnras, 465, 1008, \dodoi{10.1093/mnras/stw2748}

\bibitem[{{Villebrun} {et~al.}(2019){Villebrun}, {Alecian}, {Hussain},
  {Bouvier}, {Folsom}, {Lebreton}, {Amard}, {Charbonnel}, {Gallet},
  {Haemmerl{\'e}}, {B{\"o}hm}, {Johns-Krull}, {Kochukhov}, {Marsden}, {Morin},
  \& {Petit}}]{villebrun2019}
{Villebrun}, F., {Alecian}, E., {Hussain}, G., {et~al.} 2019, \aap, 622, A72,
  \dodoi{10.1051/0004-6361/201833545}

\bibitem[{{Wyatt}(2003)}]{wyatt2003}
{Wyatt}, M.~C. 2003, \apj, 598, 1321, \dodoi{10.1086/379064}

\bibitem[{{Younes} {et~al.}(2017){Younes}, {Baring}, {Kouveliotou}, {Harding},
  {Donovan}, {G{\"o}{\u{g}}{\"u}{\c{s}}}, {Kaspi}, \& {Granot}}]{younes2017}
{Younes}, G., {Baring}, M.~G., {Kouveliotou}, C., {et~al.} 2017, \apj, 851, 17,
  \dodoi{10.3847/1538-4357/aa96fd}

\bibitem[{{Zimmer} {et~al.}(2000){Zimmer}, {Khurana}, \&
  {Kivelson}}]{zimmer2000}
{Zimmer}, C., {Khurana}, K.~K., \& {Kivelson}, M.~G. 2000, \icarus, 147, 329,
  \dodoi{10.1006/icar.2000.6456}

\bibitem[{{Zuckerman} {et~al.}(2010){Zuckerman}, {Melis}, {Klein}, {Koester},
  \& {Jura}}]{zuckerman2010}
{Zuckerman}, B., {Melis}, C., {Klein}, B., {Koester}, D., \& {Jura}, M. 2010,
  \apj, 722, 725, \dodoi{10.1088/0004-637X/722/1/725}

\bibitem[{{Zuckerman} \& {Reid}(1998)}]{zuckerman1998}
{Zuckerman}, B., \& {Reid}, I.~N. 1998, \apjl, 505, L143,
  \dodoi{10.1086/311608}

\end{thebibliography}

\end{document}